\documentclass[
aip,
reprint,
dcolumn,
bm,
nofootinbib,
notitlepage]{revtex4-1}
\usepackage{graphicx}
\usepackage{ulem}
\usepackage{color}
\usepackage[latin1]{inputenc}
\usepackage[T1]{fontenc}
\usepackage{amsmath}
\usepackage{amssymb}
\usepackage{braket}
\usepackage{mathrsfs}
\usepackage{hyperref}
\usepackage{upgreek}
\usepackage{pstricks-add}
\usepackage{natbib} 
\usepackage{listings}
\usepackage{tikz}
\usetikzlibrary{mindmap}

\begin{document}
\title{Non-Markovian Vibrational Relaxation Dynamics at Surfaces}
\author{Eric W. Fischer}
\affiliation{Theoretische Chemie, Institut f\"ur Chemie, Universit\"at Potsdam,
Karl-Liebknecht-Stra\ss{}e 24-25, D-14476 Potsdam-Golm, Germany}

\author{Michael Werther}
\affiliation{Institut f\"ur Theoretische Physik, Technische Universit\"at Dresden, D-01062 Dresden, Germany}

\author{Foudhil Bouakline}
\affiliation{Theoretische Chemie, Institut f\"ur Chemie, Universit\"at Potsdam,
Karl-Liebknecht-Stra\ss{}e 24-25, D-14476 Potsdam-Golm, Germany}

\author{Frank Grossmann}
\affiliation{Institut f\"ur Theoretische Physik, Technische Universit\"at Dresden, D-01062 Dresden, Germany}

\author{Peter Saalfrank}
\email{peter.saalfrank@uni-potsdam.de}
\affiliation{Theoretische Chemie, Institut f\"ur Chemie, Universit\"at Potsdam,
Karl-Liebknecht-Stra\ss{}e 24-25, D-14476 Potsdam-Golm, Germany}
\affiliation{Institut f\"ur Physik und Astronomie, Universit\"at Potsdam, Karl-Liebknecht-Stra\ss e 24-25, D-14476 Potsdam-Golm, Germany}

\let\newpage\relax

\begin{abstract}
Vibrational dynamics of adsorbates near surfaces plays both an important role for applied surface science and as model lab for studying fundamental problems of open quantum systems. We employ a previously developed model for the relaxation of a D-Si-Si bending mode at a D:Si(100)-(2$\times$1) surface, induced by a ``bath'' of more than $2000$ phonon modes [U. Lorenz, P. Saalfrank, Chem. Phys. {\bf 482}, 69 (2017)], to extend previous work along various directions. First, we use a Hierarchical Effective Mode (HEM) model [E.W. Fischer, F. Bouakline, M. Werther, P. Saalfrank, J. Chem. Phys. {\bf 153}, 064704 (2020)] to study relaxation of higher excited vibrational states than hitherto done, by solving a high-dimensional system-bath time-dependent Schr\"odinger equation (TDSE). In the HEM approach, (many) real bath modes are replaced by (much less) effective bath modes. Accordingly, we are able to examine scaling laws for vibrational relaxation lifetimes for a realistic surface science problem. Second, we compare the performance of the multilayer multiconfigurational time-dependent Hartree (ML-MCTDH) approach with the recently developed coherent-state based multi-Davydov D2 {\it ansatz} [N. Zhou, Z. Huang, J. Zhu, V. Chernyak, Y. Zhao, {J. Chem. Phys.} {\bf 143}, 014113 (2015)]. Both approaches work well, with some computational advantages for the latter in the presented context. Third, we apply open-system density matrix theory in comparison with basically ``exact'' solutions of the multi-mode TDSEs. Specifically, we use an open-system Liouville-von Neumann (LvN) equation treating vibration-phonon coupling as Markovian dissipation in Lindblad form to quantify effects beyond the Born-Markov approximation.
\end{abstract}

\let\newpage\relax
\maketitle
\newpage

\section{Introduction}
Vibrational relaxation of adsorbates near surfaces continues to be of interest both for applied surface science as well as for being a prototypical example for the dynamics of open quantum systems. The coupling of the adsorbate, in what follows often denoted as a ``system'', to surface degrees of freedom, a ``bath'', leads to dissipation of excess vibrational energy and decoherence effects. This causes various phenomena, ranging, for example, from spectral line broadening in vibrational spectroscopy\cite{GuyotSionnest1995} over inelastic scattering phenomena\cite{huang2000,kroes2016} and altered chemical reactivity at surfaces\cite{guo1999}, to quite applied aspects, {\it e.g.}, the protection of passivated semiconductor surfaces in microelectronic devices \cite{avouris,grasser}. 

In the present work, 
 coupling of an adsorbate to phonons will be illustrated by
 the specific example of phonon-driven vibrational multilevel relaxation of a D-Si-Si 
 bending mode on a fully deuterium-covered, 
reconstructed silicon surface, D:Si(100)-(2$\times$1).
 For this system, dissipation due to electron-hole pair 
 creation can be neglected (in contrast to metal surfaces, where this channel often 
 dominates \cite{saalfrank2006,arnolds2011}). Vibrational relaxation dynamics is caused by efficient one-phonon processes since the system frequency 
 lies within the phonon band of the Si surface. (In contrast, for H:Si(100)-(2$\times$1) the 
 Si-Si-H bending mode lies outside the band and at least two-phonon terms 
 need to be included \cite{andrianov2006,bouakline2017}.) For D:Si(100)-(2$\times$1), 
 based on quantum chemical calculations, a realistic model Hamiltonian has been 
devised in Ref.\cite{lorenz2017}, comprising  
  an anharmonic system mode non-linearly coupled to more than $2000$ surface oscillators (``phonons''). 
 This model has been used in previous work to study the multi-dimensional 
 quantum dynamics typical for an ``open system'' 
 \cite{bouakline2017,bouakline2019,fischer2020} 
 and will be applied here also, for additional aspects not covered previously.
 We note that the techniques to be applied below are not restricted 
to D:Si(100).

Specifically, 
 in Refs.\cite{bouakline2017,bouakline2019,fischer2020} a full system-bath time-dependent Schr\"odinger Equation (TDSE) 
 was solved for the combined system-bath problem.
 Due to the exponential scaling of the TDSE with the number of 
 system plus bath degrees of freedom, various 
 adjustments and / or approximations had to be made to make the 
 problem tractable. 
 In Ref.\cite{bouakline2017}, a so-called Bixon-Jortner 
 model \cite{bixon1,bixon2} was constructed from the original Hamiltonian
 in the space of the adsorbate-surface zeroth-order sub-Hamiltonian. 
 Starting from the lowest excited vibrational state $v_0=1$ 
 of the system mode (a D-Si-Si bending vibration), 
 the multi-dimensional TDSE was then solved to follow the relaxation dynamics 
 at $T=0$ K. 
 In Ref.\cite{bouakline2019}, a quantum mechanical ``tier model'' \cite{tier1,tier2,tier3}
 was used instead in which a vibrational basis for solving the TDSE, 
 much smaller than a brute-force generated vibrational space, 
 is constructed by a hierarchical procedure, starting from  
 some initial system-bath state. It was then possible 
 to also treat relaxation of the second excited initial vibrational state, 
 $v_0=2$. Higher $v_0$ states 
 require, approximately, a polynomial growing basis $\sim N_B^{v_0}$
 with $N_B$ being the number of bath modes.
 This is more favorable than the exponential  
 scaling of brute-force models, but still remains
 a numerical challenge. In Ref.\cite{fischer2020}, a different approach 
 to solve the system-bath TDSE was followed, 
 namely the replacement of the $N_B$ physical surface oscillators, 
 by a much smaller number, $M$, of ``effective bath modes''. This was achieved 
 by adapting the Hierarchical Effective Mode (HEM) formalism 
 \cite{gindensperger2006,gindensperger2007a,gindensperger2007b,hughes2009a,hughes2009b,burghardt2012} to our surface problem. It was found 
 in Ref.\cite{fischer2020}, that with $M=60$ effective 
 modes the ``exact'' relaxation dynamics of 
 the D:Si(100)-(2$\times$1) system-bath 
 model (with $N_B>2000$) could accurately and 
 efficiently be reproduced 
 up to about 2 ps. 
 Again, initial states  $v_0=1$ and $v_0=2$ were considered, 
 with timescales for vibrational relaxation being 
 in the sub-ps regime in these cases.
 In that reference, the $(M+1)$-dimensional HEM-TDSE was solved 
 with the help of  the Multiconfigurational Time-Dependent Hartree (MCTDH)\cite{meyer1990,manthe1992,beck2000} method and its multilayer extension (ML-MCTDH)\cite{wang2003,manthe2008,vendrell2011,wang2015}. 
 Although the HEM-ML-MCTDH approach was found to 
 also improve the steep scaling problem 
 with initial state excitation $v_0$ seen in 
 Ref.\cite{bouakline2019}, higher
 $v_0>2$ were not considered in Ref.\cite{fischer2020}. It is a first  goal 
 of the present work to also consider higher excited initial states.

In general, to mitigate the ``curse of dimensionality'' problem arising for 
 multi-dimensional system-bath TDSEs, different 
 approaches have been suggested in the literature. Besides ML-MCTDH (possibly in combination with HEM), 
 recent extensively studied versions of the multi-Davydov {\it ansatz}\cite{zhou2015,wang2016,huang2017,hartmann2019,chen2019} should be mentioned. Both ML-MCTDH and the multi-Davydov \textit{ansatz} are fully 
 variational methods, and can thus be converged in principle 
 to solve the TDSE with a given Hamiltonian exactly. 
 A second goal of this paper is to 
 compare the performance of the so-called 
 multi-Davydov-D2 (MD2) {\it ansatz} with ML-MCTDH, 
 both for the HEM model of  D:Si(100)-(2$\times$1) from Ref.\cite{fischer2020}

There are also many other, more approximate, methods to treat high-dimensional system-bath problems. 
 On the multi-mode TDSE side, there are approximations to MCTDH, like 
 the Time-Dependent Self-Consistent Field (TDSCF) theory (used 
 previously also for the H/D:Si(100) system \cite{paramonov2007a,paramonov2007b}), the Gaussian MCTDH (G-MCTDH) method \cite{burghardt99,burghardt03}, or the 
 Local Coherent State Approximation (LCSA) \cite{martinazzo06}. The first 
 can be considered as MCTDH with a single configuration only, while 
 the latter two
approximate the bath single particle functions (SPFs) of MCTDH  by a set of
Gaussians wave packets or coherent states. Other  
 methods based on coherent states 
 are the Coupled Coherent States (CCS) method of 
 Shalashilin and Child \cite{shala2000,child2003}  
 and Shalashilin's Multi-Configuration Ehrenfest (MCE) method \cite{Sh09,ShBu08}. More recent reviews of different Gaussian based methods to solve the time-dependent Schr\"odinger equation can be found in Refs.\cite{irpc15,irpc21}

An alternative 
 way to describe system-bath problems quantum mechanically 
  makes use of reduced density matrix theory\cite{breuer2007,nitzan2014} 
 instead of multidimensional system-bath wave functions.
 Here the dynamics of a subsystem is described 
  by solving an open system Liouville-von Neumann (LvN) equation. The bath degrees of freedom are included 
 in an implicit manner {\it via} dissipative terms, 
 which makes the LvN dynamics non-unitary. 
Typically, these methods are based on the assumption 
 of weak system-bath coupling, 
 and further additional approximations. 
 Most importantly, the Markov approximation is 
 usually made as 
 in Redfield theory\cite{redfield1965} 
 or in the Lindblad dynamical semigroup approach \cite{lindblad1976,gorini1976}, with the latter also 
 employing the so-called secular approximation. In order to overcome these approximations, among 
 other methods the formally exact hierarchical equations of motion (HEOM) method\cite{tanimura1989,tanimura2006,tanimura2015} and the hierarchy of pure states (HOPS) method\cite{suess2014} have been developed.
 
In comparison to (most) 
 reduced density matrix approaches, 
 the solution of the full TDSE is cumbersome but 
 it also offers several fundamental advantages.
  For instance, the reduced subsystem density matrix extracted from the full system-bath wave function explicitly contains information on the non-Markovian dynamics.
 The interaction of subsystem and bath needs not to be ``weak''
 and can be described by arbitrary spectral densities for the bath, while 
  reduced approaches often rely on model spectral densities.
 In this context the question arises how reliable typical approximations 
 made in reduced density matrix theory are. In particular, 
 certain measures for non-Markovianity have been 
 introduced in the literature, for instance, measures based 
 on the so-called trace distance \cite{tracenorm1,tracenorm2}. 
 More recently, these measures have been applied to multi-dimensional TDSE solutions of 
 vibration-phonon problems with an Ohmic spectral bath \cite{lorenz}, a super-Ohmic spectral bath\cite{liu2015} and 
 of a spin-boson model \cite{thoss_spin_boson}. Also other measures 
 of non-Markovianity have been suggested, based, for example, 
 on entanglement and / or system entropy\cite{Plenio07,Rivas10b,bouakline12} and on qualitative  changes of long-time behavior in  driven dissipative systems\cite{werther2020a}.

It is a third goal of the present contribution to compare the non-Markovian dynamics 
 of the full, coupled D:Si(100)-(2$\times$1) model, which contains 
 anharmonicity and realistic spectral densities to Markovian 
 (Lindblad) dynamics, with the help of various measures.

The paper is organized as follows: In Sec.\ref{theory}, 
 we recapitulate the system-bath model 
 used in this work (Sec.\ref{theory}A) and 
 summarize the various methods 
 to describe vibrational relaxation in this system.  
 Specifically, we describe the HEM model (Sec.\ref{theory}B), 
 and high-dimensional wave function methods to solve the 
 system-bath TDSE, namely, the 
 ML-MCTDH and the multi-Davydov-D2 methods (Sec.\ref{theory}C).
 A Markovian, open-system density matrix theory, based 
 on a simplified Lindblad approach to solve the LvN equation, 
 is described in Sec.\ref{theory}D.
 In Sec.\ref{sec.results}A, following our first goal mentioned 
 above,  we 
 present results 
 for the HEM-ML-MCTDH model of 
 vibrational relaxation in D:Si(100)-(2$\times$1) 
 starting from different initial states $v_0$ (of a D-Si-Si bending mode), up to 
 $v_0=5$.
 We discuss the scaling behaviour of computational 
 effort for the problem at hand, and the scaling of 
 vibrational lifetimes as well as the behaviour of coherences 
 between system vibrational levels, as a function of initial excitation, 
 $v_0$. As for the second goal, in 
  Sec.\ref{sec.results}B the performances of the 
 HEM-ML-MCTDH and the HEM-multi-Davydov-D2 methods are 
 compared.
 The third goal is addressed in Sec.\ref{sec.results}C, where 
 the HEM-ML-MCTDH wave function and the 
 Markov-Lindblad reduced density matrix approaches 
 are contrasted with each other. From the multi-dimensional system-bath wave function, a 
 reduced system density matrix is constructed. 
 We compare vibrational lifetimes and various measures 
 of non-Markovianity, namely purity, system entropy and energy flow between 
 ``system'' and ``bath'', obtained with both models. 
 Sec.\ref{sec.conclusion} summarizes and concludes this work, 
 and several Appendices specify details not covered in the main text. 
\section{Model and Methods}
\label{theory}
\subsection{The D:Si(100) Adsorbate-Surface Model}
Here, we briefly review the main aspects of the quantum mechanical / molecular mechanics (QM/MM) embedded cluster model\cite{lorenz2017} used in this work and in Refs.\cite{lorenz2017,bouakline2017,bouakline2019,fischer2020}.

In this model, a single adsorbate bending mode is treated on the level of hybrid density functional theory employing the B3LYP hybrid functional and a Si$_{70}$D$_{54}$ ``small'' cluster model (the QM part).  For this cluster, a normal mode analysis was performed and for one of the D-Si-Si modes, called ``$\perp,a$'' (perpendicular, asymmetric) in Ref.\cite{lorenz2017}, an anharmonic potential $V(q)$ was computed along the corresponding normal mode coordinate, $q$.
We then solved the corresponding stationary vibrational Schr\"odinger equation on a set of discrete grid points to obtain the vibrational eigenpairs $\{\varepsilon_v, \ket{v}\}$. The zero point energy of this mode is $228\,\text{cm}^{-1}$, and the energies up to $v=5$ are shown in Tab.\ref{tab1}. From there it is seen that the fundamental frequency is $\omega_{10}=458\,\text{cm}^{-1}$, and the next-nearest vibrational level spacings $\omega_{v v-1}$ are $\omega_{21}=459\,\text{cm}^{-1}$, $\omega_{32}=461\,\text{cm}^{-1}$, $\omega_{43}=463\,\text{cm}^{-1}$, and $\omega_{54}=466\,\text{cm}^{-1}$. This indicates overall weak anharmonicity with slightly increasing level spacings.
\begin{table}
\caption{
Various quantities characterizing the vibrational relaxation process studied in this work: Energies (relative to the ground state energy) of the first five excited system states, $\varepsilon_{v}-\varepsilon_0$, Fermi's Golden Rule (FGR) relaxation rates $\gamma_{v}$, FGR half-lifetimes $\ln(2)\,\gamma_{v}^{-1}$, half-lifetimes obtained from solving the system-bath TDSE with HEM-ML-MCTDH, $\mathrm{T}_{1/2}$, and half-lifetimes obtained from solving the Lindblad-LvN equation, ${\cal{T}}_{1/2}$.
}
\vfill
\begin{tabular}{|c|ccccc|}
\hline
$v$ & $\varepsilon_{v}-\varepsilon_0$ &
 $\gamma_{v}$ & $\ln(2)\,\gamma_{v}^{-1}$ & $\mathrm{T}_{1/2}$ & ${\cal{T}}_{1/2}$
\\
 & (cm$^{-1}$) & (fs$^{-1}$) & (fs) & (fs) & (fs) \\
\hline
1 & 458           & 0.00578 & 120 & 145 & 120 \\
2 & 917           & 0.01160 & 60  & 102 &  60 \\
3 & 1379          & 0.01759 & 39  &  79 &  39 \\
4 & 1842          & 0.02376 & 29  &  51 &  29 \\
5 & 2308          & 0.03013 & 23  &  41 &  23
\\
\hline
\end{tabular}
\label{tab1}
\end{table}

The substrate vibrations (``phonons'') were described by a force-field of the Brenner type\cite{andrianov2006} (the MM part) applied to a ``large'' Si$_{602}$D$_{230}$ cluster. A constrained normal-mode analysis for the large cluster gives $832 \times 3-7= 2489$ vibrational normal modes (one system mode projected out). Among those, there are $N_B = 2259$ substrate vibrations or other D-Si-Si bending modes with frequencies in the range $\omega_k\in[9.8,534]\,\text{cm}^{-1}$, where the upper value (534 cm$^{-1}$) is the Debye frequency of silicon. These 2259 modes are the phonon-bath modes considered in this work. 

\begin{figure}[hbt]
\begin{center}
\includegraphics[scale=1.0]{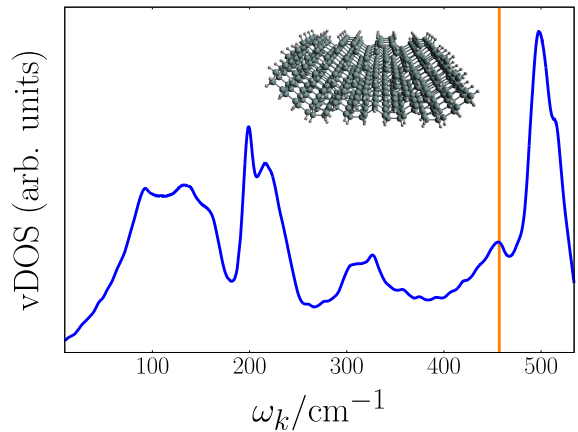}
\end{center}
\renewcommand{\baselinestretch}{1.}
\caption{Bath vibrational density-of-states (vDOS) up to the Debye frequency ($534\,\text{cm}^{-1}$) of silicon with the D-Si-Si mode fundamental vibrational transition frequency $\hbar\omega_{10}=458\,\text{cm}^{-1}$ indicated by a vertical line. The ``large'' Si$_{602}$D$_{230}$ cluster model of a D-covered Si(100)-(2$\times$1) surface is shown as an inset. The vDOS was obtained by broadening the discrete spectrum obtained from normal mode analysis, $\left\{\omega_k\right\}$, with Lorentzians of widths 10 cm$^{-1}$.
}
\label{fig.bath_vDOS}
\end{figure}

The adsorbate vibration with $\omega_{10}=458\,\text{cm}^{-1}$ lies within the silicon phonon band as indicated 
 in Fig.\ref{fig.bath_vDOS}. In consequence this mode can efficiently 
 couple to one-phonon (de-)excitations of the 
 bath, which is described by a set of bath frequencies 
 $\left\{\omega_k\right\}$ and system-bath coupling 
 constants $\left\{c_k\right\}$. The latter 
 arise from 
 a bilinear coupling approximation to the interaction Hamiltonian as in Eq.\eqref{eq.system_bath_interaction} below, 
 and were obtained from a first order phononic-expansion in the bath coordinates and a linear approximation to the coupling functions originally 
 non-linear in the system coordinate, respectively, as detailed 
 elsewhere \cite{lorenz2017,bouakline2017,bouakline2019,fischer2020}.
 In particular, a pictorial representation of the 
 coupling coefficients $\left\{c_k\right\}$ can be found 
 in Fig.2 of Ref.\cite{fischer2020}.
\subsection{The System-Bath Hamiltonian}
As in previous work\cite{lorenz2017,bouakline2017,bouakline2019,fischer2020}, the adsorbate-surface problem is described by a system-bath Hamiltonian of the form 
\begin{equation}
\hat{H}
=
\hat{H}_S
+
\hat{H}_I
+
\hat{H}_B
\quad.
\label{eq.system_bath_hamiltonian}
\end{equation}
We group the system contribution, $\hat{H}_S$, and the bath contribution, $\hat{H}_B$, into a zero-order Hamiltonian\cite{bouakline2019,fischer2020} 
\begin{equation}
\hat{H}_0
= \hat{H}_S + \hat{H}_B = 
\sum^{N_S-1}_{v=0}
\varepsilon_v
\ket{v}\bra{v}
+
\sum^{N_B}_{k=1}
\hbar\omega_k
\left(
\hat{b}^\dagger_k
\hat{b}_k
+
\dfrac{1}{2}
\right)
\quad.
\label{eq.zero_order_hamiltonian}
\end{equation}
The first term corresponds to the eigenstate representation of the system Hamiltonian, $\hat{H}_S$, where we consider $N_S$ vibrational eigenstates $\ket{v}$ with energies $\varepsilon_v$. The second term resembles $N_B$ harmonic bath modes with harmonic frequency $\omega_k$ for the $k^\mathrm{th}$-bath mode and phonon creation/annihilation operators $\hat{b}^\dagger_k$ and $\hat{b}_k$, respectively. Further, the system-bath interaction Hamiltonian is assumed to be bilinear and takes the form\cite{bouakline2017,bouakline2019,fischer2020} 
\begin{equation}
\hat{H}_I
=
\sum^{N_S-1}_{v>v^\prime}
q_{vv^\prime}
\left(
\hat{P}^\dagger_{vv^\prime}
+
\hat{P}_{vv^\prime}
\right)
\sum^{N_B}_{k=1}
\dfrac{c_k}{\sqrt{2}}
\left(
\hat{b}^\dagger_k
+
\hat{b}_k
\right)
\quad .
\label{eq.system_bath_interaction}
\end{equation}
Here, $\hat{P}^\dagger_{vv^\prime}=|v\rangle \langle v'|$ and $\hat{P}_{vv^\prime}=|v'\rangle \langle v|$ (with $v>v'$) are system raising/lowering operators. Further, $c_k$ are coupling coefficients (denoted $c_k'$ in Ref.\cite{fischer2020}, whereas $c_k=c_k'/\sqrt{2}$ in that reference), and $q_{vv^\prime}=\langle v|q|v'\rangle_q$ are  system transition matrix elements. The bilinear interaction Hamiltonian has two different contributions: (i) $\hat{P}^\dagger_{vv^\prime}\,\hat{b}_k$ and $\hat{P}_{vv^\prime}\,\hat{b}^\dagger_k$, which lead to energy transfer between the system and the bath, as well as (ii) $\hat{P}^\dagger_{vv^\prime}\,\hat{b}^\dagger_k$ and $\hat{P}_{vv^\prime}\,\hat{b}_k$, which simultaneously excite or deexcite 
 system and bath degrees of freedom {\it via} energy stored in the interaction term. The former terms respect the rotating wave approximation (RWA) in contrast to the latter post-RWA terms. Although the post-RWA terms are particularly relevant beyond the weak system-bath coupling limit, it turns out that they alter the initial relaxation dynamics even in the weak coupling scenario studied in this work.

At $T=0\,\mathrm{K}$, the dynamics of the adsorbate-surface system is governed by the time-dependent Schr\"odinger equation
\begin{equation}
i \hbar\dfrac{\partial}{\partial t}
\ket{\Psi_{\text{SB}}(t)}
=
\left(
\hat{H}_S
+
\hat{H}_I
+
\hat{H}_B
\right)
\ket{\Psi_{\text{SB}}(t)}
\quad,
\label{eq.sb_tdse}
\end{equation}
for the system-bath wave function $\ket{\Psi_{\text{SB}}(t)}$, initially taken as
\begin{equation}
\ket{\Psi_{\text{SB}}(t_0)}
=
\ket{v_0}\ket{0_1,\dots,0_{N_B}}
=
\ket{v_0}\ket{\underline{0}_B}
\quad.
\label{eq.initial_state_sys_bath}
\end{equation} 
Here, $\ket{v_0}$ is an initial adsorbate vibrational eigenstate with vibrational quantum number $v_0$ and $\ket{\underline{0}_B}$ is the harmonic multi-mode bath ground state. From the system-bath wave function, 
 a density operator $\hat{\rho}(t)= |\Psi_{\text{SB}}(t)\rangle \langle \Psi_{\text{SB}}(t)| $ can be constructed, whose trace over bath modes defines a reduced system density operator, 
\begin{equation}
\hat{\rho}_\text{S}(t) 
= 
\text{tr}_{B} \left\{ \hat{\rho}(t) \right\}
\quad,
\label{rhos}
\end{equation}
which can be compared to that obtained by reduced dynamics solving a LvN equation (see below).
\subsection{The Hierarchical Effective Mode Representation}
The solution of the TDSE \eqref{eq.sb_tdse} for the full system-bath problem suffers from the exponential scaling of the bath Hilbert space, due to the mentioned ``curse of dimensionality''.
 To mitigate this issue, we employ as a unitarily equivalent representation of the system-bath Hamiltonian in Eq.\eqref{eq.system_bath_hamiltonian}, the hierarchical effective mode (HEM) representation\cite{gindensperger2007a,hughes2009a,fischer2020}. The HEM representation 
 transforms the initial ``star-like'' configuration (where the system is coupled directly to $N_B$ bath modes) to a ``chain-like'' configuration 
 (in which the system couples to a first effective bath mode, the first 
 effective bath mode to a second one, the second one to a third one and 
 so forth). In the HEM model, the
 total Hamiltonian takes the form\cite{hughes2009a,fischer2020}
\begin{equation}
\hat{H}
=
\hat{H}_S
+
\sum^{M}_{m=1}
\hat{h}^{(m)}_\mathrm{eff}
+
\hat{H}^{(M)}_R
\quad,
\label{eq.hem_hamiltonian}
\end{equation}
with unaltered system Hamiltonian, $\hat{H}_S$, a collection of $M\ll N_B$ effective single-mode Hamiltonians, $\hat{h}^{(m)}_\mathrm{eff}$, and a residual bath contribution, $\hat{H}^{(M)}_R$. The first effective mode contribution is given by\cite{fischer2020}
\begin{multline}
\hat{h}^{(1)}_{\text{eff}}
=
\displaystyle\sum_{v>v^\prime}^{N_S-1}
q_{vv^\prime}\,
\dfrac{\bar{C}_0}{\sqrt{2}}
\left(
\hat{P}^\dagger_{vv^\prime}
+
\hat{P}_{vv^\prime}
\right)
\left(
\hat{B}^{\dagger}_1
+
\hat{B}_1\right)
\vspace{0.2cm}
\\
+
\hbar\Omega_1
\left(
\hat{B}^{\dagger}_1
\hat{B}_1
+
\dfrac{1}{2}
\right)
\quad .
\label{eq.1st_order_effmode}
\end{multline}
 Here, $\Omega_1$ is the harmonic frequency of the 
 first effective bath mode, calculated from 
 expressions given in Ref.\cite{fischer2020}.
 This mode 
 is coupled to the system mode by 
 a coupling coefficient $\bar{C}_0 = \sqrt{\sum_{k=1}^{N_B} c_k^2}$ with 
 $\hat{B}^{(\dagger)}_1=\sum_{k=1}^{N_B}
 \frac{c_k}{\bar{C}_0}\,\hat{b}^{(\dagger)}_k$ being 
 the first effective mode phonon creation/annihilation operators.
 The remaining ($M-1$) effective modes resemble a chain of bilinearly coupled effective bath modes with next-neighbor interactions
\begin{multline}
\hat{h}^{(m)}_{\text{eff}}
=
\bar{C}_{m-1}
\left(
\hat{B}^{\dagger}_{m-1}
\hat{B}_m
+
\hat{B}_{m-1}
\hat{B}^{\dagger}_m
\right)
\\
+
\hbar\Omega_m
\left(
\hat{B}^{\dagger}_m
\hat{B}_m
+
\dfrac{1}{2}
\right)
\quad,
\label{eq.ith_order_effmode}
\end{multline}
for $m\geq2$ with coupling coefficients $\bar{C}_{m-1}$ and harmonic frequencies $\Omega_m$.
 Expressions for $\bar{C}_{m-1}$, $\Omega_m$  and 
 and the $\hat{B}^{(\dagger)}_{m(-1)}$ can be found 
 in Ref.\cite{fischer2020}. 
 The residual bath Hamiltonian 
 $\hat{H}^{(M)}_R$ collects all ($N_B-M$) residual effective modes
 with frequencies $\Omega_j$, 
\begin{multline}
\hat{H}^{(M)}_R
=
\displaystyle\sum_{j=M+1}^{N_B}
\hbar\,d_{Mj}
\left(
\hat{B}^{\dagger}_{M}
\hat{B}_j
+
\hat{B}_{M}
\hat{B}^{\dagger}_j
\right)
\\
+
\displaystyle\sum_{j={M}+1}^{N_B}\,
\hbar\Omega_j
\left(
\hat{B}^{\dagger}_j
\hat{B}_j
+
\dfrac{1}{2}
\right)
\quad,
\label{eq.Lth_order_resbath}
\end{multline}
which couple exclusively to the $M^\mathrm{th}$ effective mode on the chain 
 {\it via} coupling coefficients $d_{Mj}$ 
 (also defined in Ref.\cite{fischer2020}).
 The main advantage of the HEM representation 
 is that it cannot 
 only be used as an exact method, 
 but also as an approximate one, 
 by truncating the Hamiltonian \eqref{eq.hem_hamiltonian} at a certain order $M$ 
 and neglecting the residual bath $\hat{H}^{(M)}_R$. 
 This is equivalent 
 to recovering the exact dynamics of the system-bath problem up 
 to a finite time\cite{gindensperger2007a,fischer2020}.
  The latter can be systematically enlarged by increasing the truncation order,
  $M$. The reduction to $M\ll N_B$ effective 
 bath modes significantly improves the 
 scaling issue of system-bath problems, 
 at least for system-bath Hamiltonians used in this work.
\subsection{System-Bath Wave Function Dynamics}
The solution of the ($M+1$)-dimensional TDSE 
 arising from the HEM is a challenging task, as 
 $M$ can still be large, depending 
 on the problem and required ``cutoff time''.
 Here, we employ the multilayer multiconfigurational time-dependent Hartree 
 approach (ML-MCTDH)\cite{wang2003} and compare its 
 performance 
 with the recently developed coherent-state based multi-Davydov-D2 approach\cite{zhou2015}. Both rely on the Dirac-Frenkel variational principle (DFVP) 
 to find an optimal system-bath wave function, $\ket{\Psi_{\text{SB}}(t)}$, 
 applying, however, different {\it ans\"atze} 
 for $\ket{\Psi_{\text{SB}}(t)}$.
\subsubsection{The Multilayer MCTDH {\it Ansatz}}
For the MCTDH {\it ansatz}\cite{wang2003,manthe2008,vendrell2011,wang2015}
 of a system-bath problem with a single system DoF and $M$ effective 
 bath modes, we first of all group the 
 latter into $m$ groups of ``combined bath modes'', 
  with $m<M$. The MCTDH system-bath wave function is then given by
\begin{equation}
\ket{\Psi_{\text{SB}}^{\text{MCTDH}}(t)}
=
\displaystyle\sum_{j_0=1}^{n_s}
\displaystyle\sum_{j_1,\dots,j_m}^{n_1,\dots,n_m}
A^{(1)}_{j_0 j_1\dots j_m}(t)
\prod^m_{\kappa=0}
\ket{\varphi^{(1;\kappa)}_{j_\kappa}(t)}
\quad ,
\label{eq.sb_mctdh_wf}
\end{equation}
with coefficients, $A^{(1)}_{j_0 j_1\dots j_m}(t)$, and orthonormal time-dependent single-particle functions (SPFs) of the system, $\ket{\varphi^{(1;s)}_{j_0}(t)}$, and combined bath modes, $\ket{\varphi^{(1;\kappa)}_{j_\kappa}(t)}$. Here, $n_s$ is the number of system SPFs and $n_\kappa$ the number of multi-mode bath SPFs for the $\kappa^\mathrm{th}$-combined mode. In the ML-MCTDH approach, the multi-mode bath SPFs are subsequently expanded in a new basis of time-dependent SPFs, 
\begin{equation}
\ket{\varphi^{(1;\kappa)}_{j_\kappa}(t)}
=
\sum^{N_1\dots N_{d_\kappa}}_{i_1,\dots,i_{d_\kappa}}
A^{(2;j_\kappa)}_{i_1\dots i_{d_\kappa}}(t)
\prod^{d_\kappa}_{\kappa^\prime=1}
\ket{\varphi^{(2;j_\kappa)}_{i_{\kappa^\prime}}(t)}
\quad,
\label{eq.multimode_spfs_primitives}
\end{equation}
which provide another layer of the ML-wave function. The iterative expansion of multi-mode SPFs in a new basis of SPFs subsequently adds new layers to the ML-wave function and is truncated by a time-independent primitive basis, which is here given by bosonic number states for a harmonic oscillator bath. 

A particular useful representation of a ML-wave function, which allows to depict its complexity, is given by the diagrammatic ``ML-trees''\cite{manthe2008}. In the present work, $M=60$ was found to converge our HEM results up to a truncation time of 2 ps, which was sufficient to describe the vibrational relaxation process. The ML-trees corresponding to $M=60$ are shown in Appendix A, where further numerical details of our HEM-ML-MCTDH treatment of the problem at hand are given. We just mention here, that we employed a spectral representation for the system in terms of its respective eigenstates and treated the effective mode bath in second quantization representation (SQR). For the system Hamiltonian, $N_S=10$ vibrational system eigenstates were taken into account, leading to $n_s=10$ SPFs in the ML-wave function. For all ML-MCTDH calculations, we employ the Heidelberg MCTDH package, version 8.6.\cite{vendrell2011,heidelbergmctdh}.
\subsubsection{The Multi-Davydov-D2 {\it Ansatz}}
In the multi-Davydov-D2 {\it ansatz}, the wave function is written as\cite{zhou2015,werther2020b,zhao2021}
\begin{align}
\ket{\Psi_{\text{SB}}^{\text{D$_2$}}(t)}
&=
\sum^{K}_{j=1}
\left(
\sum^{N_S-1}_{v=0}
A_{vj}(t)
\ket{v}
\right)
\ket{\boldsymbol{\alpha}_{j}(t)}
\quad,
\label{eq.md2_wf}
\end{align}
with time-dependent coefficients, $A_{vj}(t)$, system eigenstates, $\ket{v}$, and $K$ multi-mode coherent states (CS)
\begin{equation}
\ket{\boldsymbol{\alpha}_{j}(t)}
=
\bigotimes^M_{\kappa=1}
\ket{\alpha_{j\kappa}(t)}
\quad,
\label{eq.multi_mode_cs}
\end{equation}
modeling the bath contributions to the system-bath wave function. $\ket{\boldsymbol{\alpha}_{j}(t)}$ is composed as a direct product of $M$ normalized, single-mode CS given by
\begin{equation}
\ket{\alpha_{j\kappa}(t)}
=
\exp\left(-\dfrac{1}{2}\vert\alpha_{j\kappa}(t)\vert^2\right)
\exp\biggl(\alpha_{j\kappa}(t)\,\hat{B}^\dagger_\kappa\biggr)
\ket{0_\kappa}
\quad,
\label{eq.one_dim_cs}
\end{equation}
{with complex, time-dependent displacements, $\alpha_{j\kappa}(t)$. The real part of these displacements is proportional to the position and the imaginary part is proportional to the momentum of the $j^\mathrm{th}$-bath oscillator. The time-evolution of the coefficients $A_{vj}(t)$ and the complex displacements $\alpha_{j\kappa}(t)$ are
  governed by the DFVP. In the limit of a large multiplicity $K$, 
 which is a convergence parameter, 
 the exact wave function is recovered. Details about the numerical implementation, which shares some common aspects, like regularization, with ML-MCTDH are given 
 in Ref.\cite{werther2020b}.}

Recently, the multi-Davydov \textit{ansatz} methodology has been used in a benchmark calculation of the sub-Ohmic spin-boson model, comparing with results from the HOPS method as well as ML-MCTDH \cite{hartmann2019}. 

{Like MCTDH, the multi-Davydov 
 method is fully variational. Indeed for both methods the expansion coefficients as well as the wave function parameters are determined fully variationally.
 This distinguishes the multi-Davydov approach from other 
 CS-based methods, like the multi-configuration Ehrenfest method \cite{Sh09}, which is an extension to multi-surface dynamics of the classical trajectory-based coupled coherent states method, discussed in Ref.\cite{ShBu08}. This fact allows for a very favorable scaling of the multiplicity $K$ (and thus the numerical effort of the calculation) with the number of degrees of freedom, to be discussed in more detail below. 
 However, like MCTDH, this comes at the price of the loss of the relative simplicity of the equations of motion, 
 in the Davydov case for the 
  coherent state parameters, which fulfill nonclassical, highly nonlinear and coupled equations of motion. 

\subsection{Reduced System Quantum Dynamics}
A conceptually different approach to system-bath quantum dynamics relative to the full TDSE approach relies on the direct propagation of the system reduced density matrix, $\hat{\rho}_S(t)$, which fully governs the system dynamics\cite{breuer2007,nitzan2014}, $\hat{\rho}_S(t)$. The latter can be obtained from the full system-bath wave function as illustrated in Eq.(\ref{rhos}) above. In the system eigenbasis $\{\ket{v}\}$, the diagonal elements, $\rho_{vv}(t)$, of $\hat{\rho}_S(t)$ are the populations of the vibrational states $\ket{v}$, and the off-diagonal elements $\rho_{vv^\prime}(t)$ are known as vibrational coherences. Here, we concentrate on the system populations and vibrational coherences of energetically adjacent system states, $\rho_{vv+1}(t)$, respectively.

A prominent example is the open-system Liouville-von Neumann equation in Lindblad form\cite{lindblad1976,gorini1976}. Here we use a LvN-Lindblad equation for $T=0\,\mathrm{K}$ of the form 
\begin{equation}
\dfrac{\partial}{\partial t} \hat{\rho}_S
=
-
\dfrac{\text{i}}{\hbar}
[\hat{H}_S,\hat{\rho}_S]
+
\displaystyle\sum_v
\gamma_v
\left(
\hat{C}_v\hat{\rho}_S\hat{C}^{\dagger}_v
-
\dfrac{1}{2}\left[\hat{C}^{\dagger}_v\hat{C}_v,\hat{\rho}_S\right]_+
\right) .
\label{eq.liouville_von_Neumann_Lindblad}
\end{equation}
This equation arises from a more general Lindblad equation 
 by assuming that only transitions $v \rightarrow v-1$ 
 are triggered by the environment, as outlined and justified 
 in Appendix B. 
 In Eq.(\ref{eq.liouville_von_Neumann_Lindblad}), the first term 
 on the r.h.s. describes the unitary evolution of the 
 system density operator, and the 
  second one its non-unitary evolution due to 
 a Markovian, dissipative Lindblad- Liouvillian which accounts for the coupling of the 
 system to the bath.
 In this term, 
 Lindblad operators $\sqrt{\gamma_v}\,\hat{C}_v$ appear, where $\gamma_v$ is the transition rate 
 from state $|v\rangle$ to state $|v-1\rangle$, 
 and $\hat{C}_v=|v-1\rangle \langle v|$ 
 a ladder operator which describes this transition. 
 $[\cdot,\cdot]_+$ denotes an anticommutator.
 The rates $\gamma_v$ are calculated from 
 Fermi's Golden Rule (FGR) using our microscopic system-bath model 
 for D:Si(100), as outlined in detail in Appendix B. 
 The lowest five rates $\gamma_1$ to $\gamma_5$ are listed 
 in Tab.\ref{tab1}, ranging from about (173 fs)$^{-1}$ for 
$v=1$ to about (30 fs)$^{-1}$ for
$v=5$.
 Eq.(\ref{eq.liouville_von_Neumann_Lindblad}) is solved subject to 
 the initial condition
\begin{equation}
\hat{\rho}_S(t=0)
=
|v_0\rangle \langle v_0|
\quad,
\end{equation} 
with $v_0=1,2, \dots, 5$ in what follows. The dissipative part causes vibrational population transfer (changing of $\rho_{vv}$) and dephasing (decaying $\rho_{vv'}$). In the Lindblad model, the equations of motion of the reduced density matrix in the system eigenstate-representation are given by
\begin{eqnarray}
  \frac{d \rho_{vv}}{dt} & = & - \gamma_v \ \rho_{vv} +  
 \gamma_{v+1}  \ \rho_{v+1 v+1}\quad,\label{lind1} \\
  \frac{d \rho_{vv'}}{dt} & = & \rho_{vv'} \left[-i \omega_{vv'} -
  \frac{1}{2} (\gamma_{v}+\gamma_{v'}) \right] , \ v \neq v' \label{lind2} ,
  \end{eqnarray}
with $\omega_{vv'}=(\varepsilon_v - \varepsilon_{v'})/\hbar$.

The advantage of Eq.\eqref{eq.liouville_von_Neumann_Lindblad} over the full system-bath TDSE \eqref{eq.sb_tdse} results from the significantly different scaling behavior with respect to the underlying Hilbert space. A basis representation of Eq.\eqref{eq.liouville_von_Neumann_Lindblad} scales only as $N^2_S$, where $N_S$ is the number of system states. This simplifies the propagation of the system-bath problem tremendously, but at the cost of a series of approximations inherent in Eq.\eqref{eq.liouville_von_Neumann_Lindblad}, among them\cite{breuer2007}: (i) Conservation of the initial state's product structure, $\hat{\rho}(t_0)=\hat{\rho}_S\otimes\hat{\rho}_B$ (Born approximation), (ii) the neglect of memory effects of the bath, \textit{i.e.}, the LvN-equation in Lindblad form is local in time (Markov approximation) and (iii) a unidirectional energy flow from the system to the bath at $T=0\,\mathrm{K}$. 

By definition, the open-system LvN-equation in Lindblad form is a Markovian quantum master equation\cite{breuer2007}. Hence, a comparison between the dynamics resulting from Eq.\eqref{eq.liouville_von_Neumann_Lindblad} and the full system-bath TDSE \eqref{eq.sb_tdse} allows to extract qualitative signatures of ``non-Markovian'' effects.

We solve the open-system LvN-equation in Lindblad form numerically by means of the QuTiP package\cite{johansson2012,johansson2013} in combination with Python 3.6.
\section{Results and Discussion}
\label{sec.results}
We begin our discussion with the reduced scaling behaviour of the truncated HEM bath basis for one-phonon driven vibrational relaxation processes. For an initial state analogous to Eq.\eqref{eq.initial_state_sys_bath}, the relaxation process can be depicted schematically by the cascade
\begin{equation}
\ket{v_0}
\to
\ket{v_0-1}
\to
\dots
\to
\ket{2}
\to 
\ket{1}
\to 
\ket{0}
\quad,
\label{eq.relaxation_cascade}
\end{equation}
where every transition is accompanied by a single-phonon excitation in the bath. We note that 
 overtone transitions like $\ket{v_0}\to\ket{v_0-2}$ are unimportant 
 here as the phonon band of our model bath does not support modes with appropriate frequencies. The number of bath mode basis states, $n_B(v_0)$, which are required to describe the single-phonon mediated relaxation has recently been shown to scale polynomially with the number of bath modes $N_B$ as\cite{bouakline2019}
\begin{equation}
n_B({v_0}) 
=
\sum^{v_0}_{k=1}
\dfrac{(N_B+k-1)!}{k!\,(N_B-1)!}
\quad,\,
v_0
=
1,2,\dots
\quad.
\label{eq.basis_scaling_onephonon}
\end{equation}
In leading order, Eq.\eqref{eq.basis_scaling_onephonon} takes the form
\begin{equation}
n_B(v_0) 
\sim
N^{v_0}_B
\quad,
\,
N_B\,\mathrm{large}
\quad,
\label{eq.leadingorder_basis_scaling_onephonon}
\end{equation}
which is the scaling behavior mentioned in the Introduction. This scaling renders a straightforward study of relaxation processes here with $N_B=2259$ already for $v_0>2$ prohibitively expensive for the ``tier model'' approach in Ref.\cite{bouakline2019}. In case of the truncated HEM representation, $N_B$ in Eq.\eqref{eq.basis_scaling_onephonon} is replaced by $M$ such that $n^{v_0}_B\sim M^{v_0}$ for sufficiently large $M$ in Eq.\eqref{eq.leadingorder_basis_scaling_onephonon}, respectively. In this work, we find the relaxation dynamics over a time-interval of $t_f=2000\,\mathrm{fs}$ to be exactly recovered with only $M=60$ effective modes for all initial states with $v_0=1$ to $v_0=5$, respectively. 
 In general, the truncation order $M$ depends on both the natural time-scale of the process under study and the nature of the  system-bath interaction. 
Furthermore, it has also to be assumed that the truncation order is in general not independent of the initial system state, in contrast to what will be discussed below for our adsorbate-surface model. 
\subsection{One-Phonon-Driven Multilevel Relaxation}
\label{subsec.multilevel_relaxation}
We now turn to the vibrational relaxation dynamics of the excited D-Si-Si-bending mode for initial vibrational quantum numbers $v_0=1$ to $v_0=5$. In Fig.\ref{fig.dynamics_mlmctdh} (top-rows), the time-evolution of the reduced system populations, $\rho_{vv}(t)$, is shown for a time interval of $t_f=2000\,\mathrm{fs}$ with $M=60$. Here the TDSE was solved with the ML-MCTDH method. 
\begin{figure*}[hbt]
\begin{center}
\includegraphics[scale=1.0]{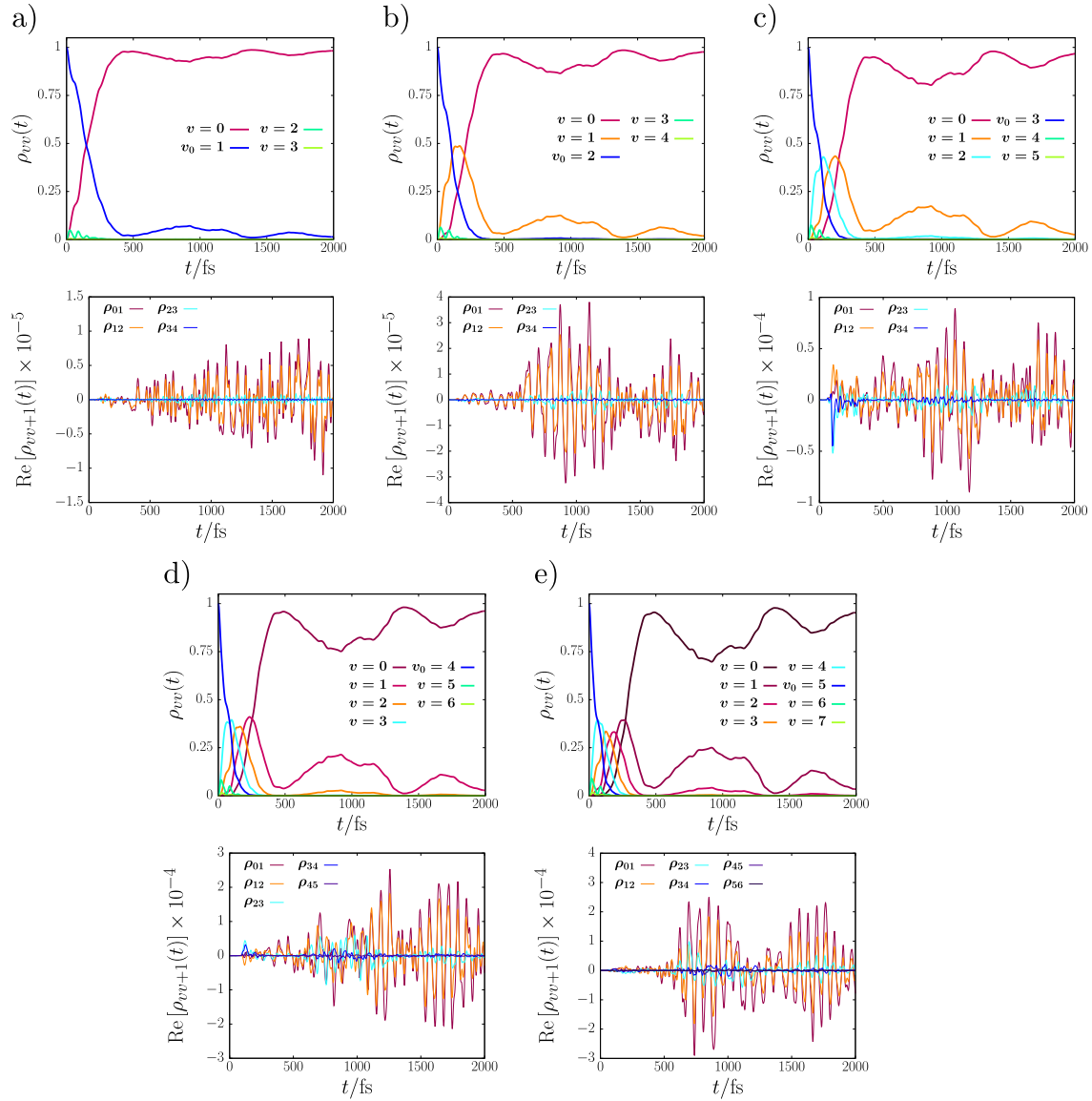}
\end{center}
\renewcommand{\baselinestretch}{1.}
\caption{Top rows: Converged time-evolution of system vibrational populations, $\rho_{vv}(t)$, for initial vibrational system states $\ket{v_0}$ with quantum numbers $v_0=1$ to $v_0=5$ a)-e), obtained from full system-bath TDSE with truncated HEM Hamiltonian for $M=60$, solved with ML-MCTDH. Bottom rows: Time-evolution for real part of corresponding vibrational coherences, $\mathrm{Re}\left[\rho_{vv+1}(t)\right]$, for energetically adjacent system states.}
\label{fig.dynamics_mlmctdh}
\end{figure*}
The relaxation dynamics takes place at a time scale of roughly $500\,\mathrm{fs}$ for all initial states studied here. For longer times, two damped recurrences in $\rho_{11}(t)$ and $\rho_{22}(t)$ can be observed around $1000\,\mathrm{fs}$ and $1750\,\mathrm{fs}$, with magnitudes increasing with the system initial vibrational state. The recurrences can be related to the structured bath vDOS, for which only a few modes dominate the initial energy transfer process. 
 This gives rise to damped Rabi-type oscillations, causing the 
 recurrences.

 Despite the non-exponential decay, it is useful to define a
 half-lifetime, $\mathrm{T}_{1/2}$, after which  
 the population of initial state $v_0$ has dropped to 
 $1/2$. From Tab.\ref{tab1} and also from 
 Fig.\ref{fig.dynamics_lindblad_vs_mlmctdh} (see below), 
 we find that the half-lifetimes 
 range from 145 fs for $v=v_0=1$ to 
 41  fs for $v=v_0=5$. For $v_0=1$ and $v_0=2$, these 
 lifetimes are in good agreement with earlier work,
 where the ``tier'' model had been used to solve the 
 TDSE for D:Si(100)-(2$\times$1) \cite{bouakline2019}. 
 We also find that the half-lifetimes shorten with 
 increasing $v_0$. In fact, it is well-known that for 
 a harmonic oscillator 
 coupled bilinearly to a harmonic bath, Fermi's Golden 
 Rule provides at $T=0$ K, strict selection rules 
 $\Delta v=-1$ and a scaling
 law\cite{bouakline12}
\begin{equation}
{T}^{(v_0)}_{1/2}
=
\dfrac{{T}^{(1)}_{1/2}}{v_0}
\quad,
\label{eq.markov_lifetime_scaling}
\end{equation}
 for the lifetimes.
 Note that this ideal scaling
 is not well 
 fulfilled for the ``exact'' solution 
 of the system-bath TDSE,  despite
 fitting trends: For instance, 
 $\mathrm{T}_{1/2}$ should be 29 fs for $v=5$ 
 according to Eq.(\ref{eq.markov_lifetime_scaling}) 
 but is 41 fs according to Tab.\ref{tab1}. 
 In contrast, the ideal scaling law is nicely 
 reproduced by the FGR half-lifetimes 
 $\ln(2)\,\gamma_v^{-1}$; only for higher 
 $v$ a small deviation from ideal scaling is seen 
 (the half-lifetime of $v=5$ should be 24 fs, not 23 fs, 
 for example), indicating the only weak anharmonicity of 
 our system. Note also that the ``exact'' lifetimes 
 are all consistently larger than the 
 FGR ones, in agreement with findings already 
 reported in Ref.\cite{bouakline2019}. 

Besides the relaxation process (\textit{cf.}~Eq.\eqref{eq.relaxation_cascade}), we additionally observe the population of vibrational states lying energetically above the initial state according to processes
\begin{equation}
\dots
\leftarrow
\ket{v_0+2}
\leftarrow
\ket{v_0+1}
\leftarrow
\ket{v_0}
\quad,
\label{eq.post_rwa_cascade}
\end{equation}
which result from the post-rotating wave contribution in the bilinear system-bath interaction in Eq.\eqref{eq.1st_order_effmode}. This process alters in particular the initial decay of the system excitation for small times up to around $100\,\mathrm{fs}$ and the corresponding states are depopulated at longer times. We note 
 that the post-RWA terms do not alter the computational 
 scaling problem substantially, as we observe only a significant population of states $\ket{v_0+1}$ and $\ket{v_0+2}$.

Complementing the population dynamics, the time-evolution of vibrational coherences, $\rho_{vv+1}(t)$, provides additional insight into the relaxation process. In Fig.\ref{fig.dynamics_mlmctdh} (bottom rows), we show the corresponding real part, $\mathrm{Re}\left[\rho_{vv+1}(t)\right]$, where small but clearly non-zero oscillatory vibrational coherences emerge roughly after $t=50\,\mathrm{fs}$, which gradually increase in magnitude and are dominated by the $\rho_{01}$ and $\rho_{12}$ contributions. 
The overall coherence dynamics is determined by ``beats'' occurring parallel to the recurrences in the populations. As the latter relates to a re-excitation of the system, it leads to a coherent contribution of several system states to the full wave function, which in turn manifests in enhanced coherence amplitudes. For times $t>1000\,\mathrm{fs}$, the coherences $\rho_{01}$ and $\rho_{12}$ oscillate at the same frequency but different amplitudes. 
The inverse frequency, {\it i.e.}, the oscillation period $T$, is about 73 fs, reflecting nicely the energy difference between system states $\ket{0}$ and $\ket{1}$ (or $\ket{1}$ and $\ket{2}$, which is similar), through $T=2 \pi/\omega_{10} \sim 2 \pi/\omega_{21}$. Interestingly, although the population is mainly determined by contributions from system eigenstates $\ket{0},\ket{1}$ and $\ket{2}$, coherences $\rho_{23}$ and $\rho_{34}$ show minor but significant contributions, which are especially pronounced during the recurrence events. Also their frequencies are about the same as for $\rho_{01}$ and $\rho_{12}$, because also the higher level spacings are similar to $\hbar \omega_{10}$. The non-exponential decay of populations and the occurrence of coherences are signatures of non-Markovian dynamics (see below).
\subsection{Multi-Davydov D2 vs. ML-MCTDH}
We now compare the multilevel relaxation dynamics obtained with ML-MCTDH and the multi-Davydov D2 approach. For both, the same HEM bath model (with $M=60$), system and system-bath coupling parameters were adopted. In Fig.\ref{fig.dynamics_md2_vs_mlmctdh}, the population dynamics is shown for both methods and three different initial states, $v_0=3$, $4$, and $5$. It can be seen that the two different approaches give identical results to within line thickness for times up to 500~fs. After that time, the results start to deviate slightly. The deviation is higher with increasing initial excitation, $v_0$. (The deviation is vanishingly small for $v_0=1$ and $v_0=2$ and therefore the resulting curves are not shown.)

Both results are converged with respect to the basis used for ML-MCTDH (see Appendix A) and the multiplicity $K$ of the multi-Davydov-D2 method, respectively, but rely on different implementations on different platforms. The multi-Davydov results have been generated on a modern laptop computer using MATLAB, whereas the ML-MCTDH results were obtained {\it via} the Heidelberg MCTDH package running on a modern workstation. This may explain the slight deviations between both methods for $v_0>3$ after the first recurrences, however, we cannot give a precise reason for these deviations at present.
}
\begin{figure}[hbt]
\begin{center}
\includegraphics[scale=1.0]{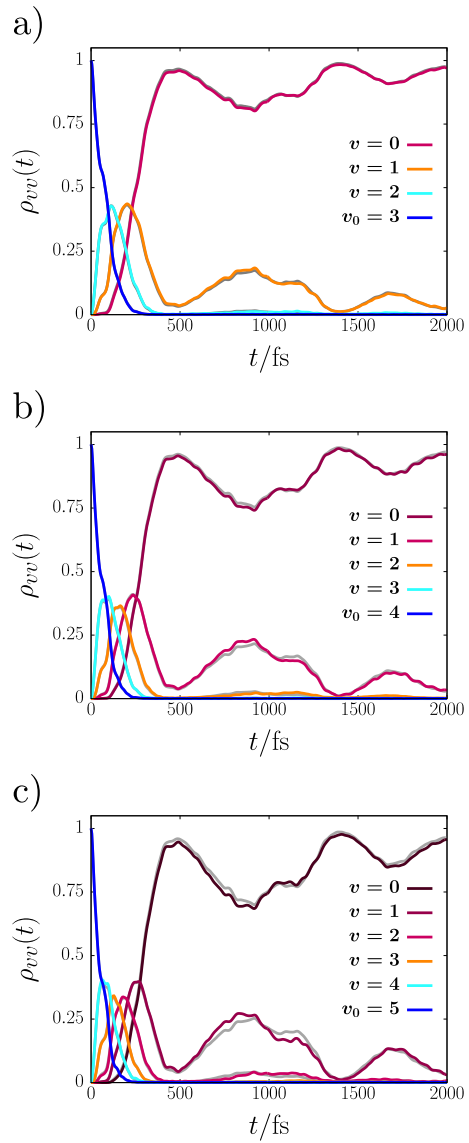}
\end{center}
\renewcommand{\baselinestretch}{1.}
\caption{Population dynamics from multi-Davydov-D2 {\it ansatz} (colored) and ML-MCTDH approach (grey) for initial vibrational system states $\ket{v_0}$ with quantum numbers (a) $v_0=3$, (b) $v_0=4$ and (c) $v_0=5$, obtained from full system-bath TDSE with truncated HEM Hamiltonian for $M=60$ effective modes and MD2 multiplicity $K=30$.}
\label{fig.dynamics_md2_vs_mlmctdh}
\end{figure}
The scaling of the numerical effort of the multi-Davydov implementation is proportional to $K^s$ with an exponent $s$ between two and three. However, surprisingly, the multiplicity needed to converge the results does not depend much on the number of degrees of freedom. We have performed calculations with 100 bath modes (not shown) and have found that the same number $K=30$ of basis functions was sufficient for convergence. It turns out that, as in Ref.\cite{hartmann2019}, the ratio K/M, \textit{i.e.}, multiplicity over number of bath modes, is (much) smaller than unity. The initial excitation seems to have a bigger influence on the numerical effort, however, which for identical $K$ went up slightly for larger values of $v_0$. The adaptive step size integrator used in the MATLAB implementation needed longer to converge the results with the required accuracy for increased values of $v_0$. A ballpark number for the absolute time needed for the run of the program that generated the multi-Davydov data set in Fig.\ref{fig.dynamics_md2_vs_mlmctdh}(c) was 15000 seconds on a modern laptop computer. In contrast, the ML-MCTDH calculations for the same parameter set took 54000 seconds. In this context, we note that for the  ML-MCTDH calculations, we employed a similar ML-tree as in Ref.\cite{fischer2020}, which was, however, not optimized with respect to computation time. Further, as convergence criterion for the number of SPFs, we consider a lowest natural population of $\leq 10^{-4}$ for every node of the ML-tree, which is relatively tight and might be weakened. 

All in all, however, the multi-Davydov {\it ansatz} as discussed here seems to be a very promising approach for system-bath problems with relatively large harmonic bath Hamiltonians as it is efficient and only the multiplicity $K$ needs to be converged. The ML-MCTDH approach is more versatile and general but also more complex, as it requires a proper topology of the ML-tree as well as convergence of primitive basis functions and SPFs. This renders the multi-Davydov a relatively straightforward approach in combination with the HEM-representation. Nevertheless, it might be instructive to study the dependence of the multiplicity $K$ in the multi-Davydov D2 approach on the number of bath modes, which has been recently indicated to behave linearly for baths up to $N_B=300$ modes\cite{fujihashi2017}. This will be a critical test also against ML-MCTDH, which was successfully applied to a spin-boson model involving several thousand harmonic modes\cite{wang2008} (without the HEM-representation of the bath Hamiltonian).
\subsection{Markovian {\it vs.} Non-Markovian Relaxation Dynamics}
\label{subsec.markovian_multilevel_relaxation}
\begin{figure*}[hbt]
\begin{center}
\includegraphics[scale=1.0]{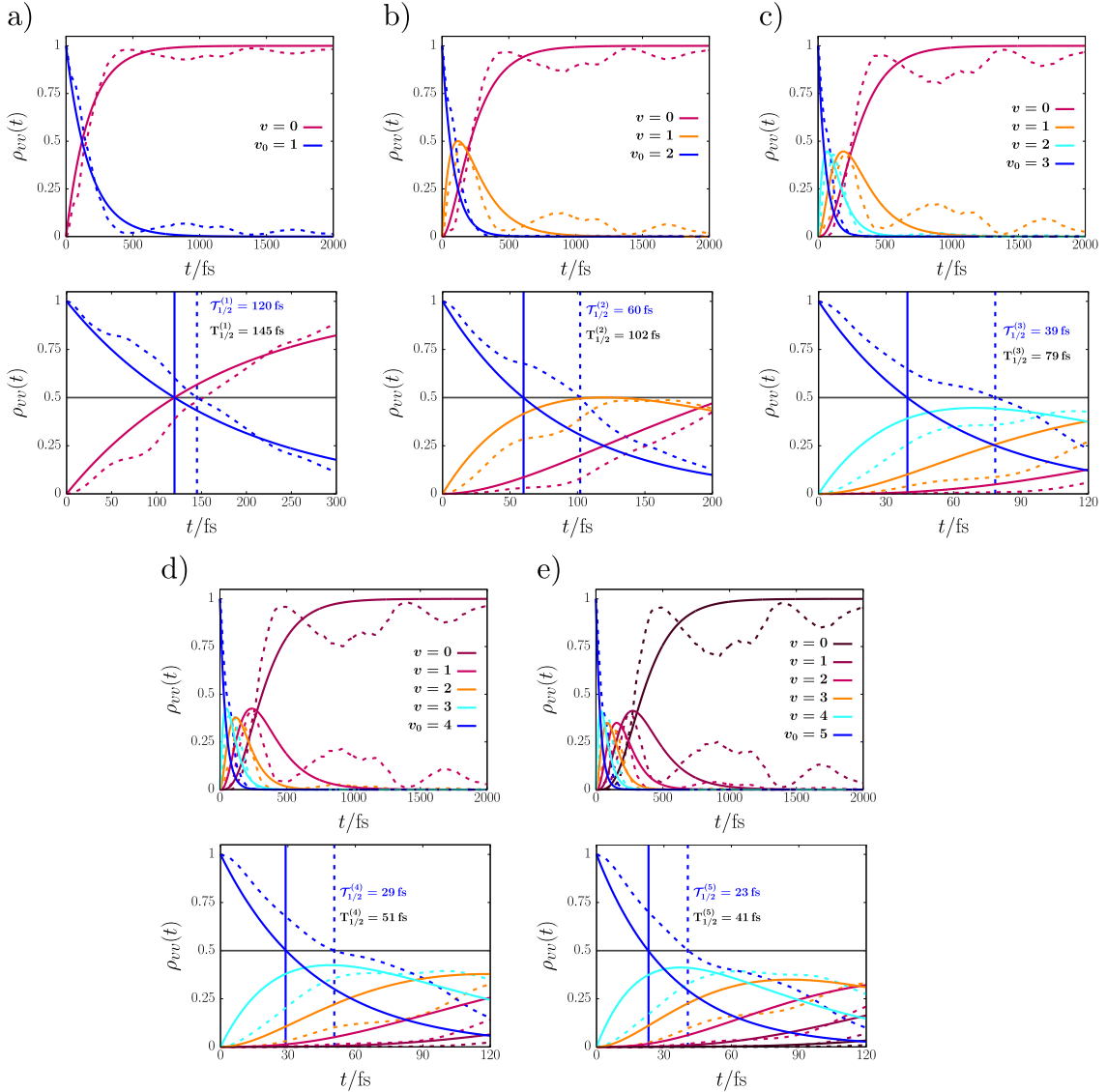}
\end{center}
\renewcommand{\baselinestretch}{1.}
\caption{Top rows: Markovian system vibrational populations (solid lines) compared to non-Markovian populations (dashed lines) for $v_0=1$ to $v_0=5$ a)-e). Bottom rows: Initial relaxation dynamics with Markovian, $\mathcal{T}^{(v_0)}_{1/2}$, and non-Markovian (truncated HEM representation) half-lifetimes, $\mathrm{T}^{(v_0)}_{1/2}$, in dependence of $v_0$.
}
\label{fig.dynamics_lindblad_vs_mlmctdh}
\end{figure*}
The full system-bath dynamics is of course non-Markovian and can be compared to Markovian, reduced dynamics obtained from the open system LvN-equation in Lindblad form \eqref{eq.liouville_von_Neumann_Lindblad}. Here we use three sets of measures to quantify non-Markovian behavior, namely the reduced-density matrix itself with diagonal (populations) and off-diagonal elements (coherences), respectively, purity and von Neumann-entropy, and, finally, the energy flow between system and bath.

\subsubsection{Populations and Coherences}
In Fig.\ref{fig.dynamics_lindblad_vs_mlmctdh}, top rows, we reconsider the population dynamics for various initial states $|v_0\rangle$. As dashed lines, we show the same populations $\rho_{vv}(t)$ as in Fig.\ref{fig.dynamics_mlmctdh}, obtained with the ``exact'' HEM-ML-MCTDH model, and compare them with Markovian system vibrational populations obtained with the Lindblad model (solid lines). In the bottom rows of Fig.\ref{fig.dynamics_lindblad_vs_mlmctdh}, the same information is given at shorter times up to 120 fs, and also half-lifetimes are indicated, obtained either from ML-MCTDH ($\mathrm{T}_{1/2}$) or from the Lindblad/FGR model (${\mathcal{T}}_{1/2}$). These half-lifetimes are also listed in Tab.\ref{tab1} as mentioned earlier.

We observe two main differences in the Markovian limit: First, the populations do not show a particular fine structure opposed to the oscillatory nature of the full system-bath scenario and, second, no recurrences are observed. These observations relate to two properties of the Lindblad approach: (i) The bath is assumed to be infinitely large with a constant vibrational density-of-states (vDOS) and (ii) the population transfer is unidirectional from the system to the bath and in particular no post-RWA effects are taken into account due to the secular approximation. As a  third observation, the Lindblad relaxation dynamics of the initial state is exponential in contrast to the highly non-exponential decay resulting from the full dynamics. 

The Lindblad / FGR half-lifetimes trivially coincide with the 
 FGR half-lifetimes ${\mathcal{T}}_{1/2}^{(v_0)}=\ln(2)\,\gamma_{v_0}^{-1}$ 
 and are longer than those of the full-TDSE, $T_{1/2}^{(v_0)}$. 
 Of course, the ${\mathcal{T}}_{1/2}^{(v_0)}$ then also fulfill 
 the ideal scaling law 
 of Eq.(\ref{eq.markov_lifetime_scaling}) better.

Another striking difference between the reduced LvN-Lindblad and the full system-bath dynamics is the absence of any coherences
 for the former, in contrast to the latter for which
 coherences were found in Fig.\ref{fig.dynamics_mlmctdh}, bottom rows. 
 In the Lindblad approach, coherences do not appear 
 because the initial state is a projector and therefore 
 all off-diagonal elements of the system density matrix 
 are zero. In particular, the latter stay zero,
 because coherences and populations are strictly 
 decoupled in the Lindblad model, {\it cf.} Eqs.\eqref{lind1}
 and \eqref{lind2}. This is not strictly 
 so in other Markovian theories, like Redfield theory\cite{redfield1965}, where  the secular approximation is not made 
 and coherence-population transfer terms exist. 

\subsubsection{Purity and von Neumann-Entropy}
\label{subsec.reduced_systemw_properties}
The purity, $p_s(t)$, and the von Neumann-entropy, $S_\mathrm{vN}(t)$, are other useful measures of non-Markovianity. The purity of the reduced system is defined as
\begin{equation}
p_s(t)
=
\text{tr}_S\{\hat{\rho}^2_S(t)\}
\leq
1
\quad,
\label{eq.purity}
\end{equation}
where the equality holds only if $\hat{\rho}_S$ corresponds to a pure state, \textit{i.e.}, here initially with $\hat{\rho}_S(0)=\ket{v_0}\bra{v_0}$. Otherwise, the reduced system is in a \textit{mixed} state. Further, we consider the von Neumann-entropy
\begin{equation}
S_{\text{vN}}(t)
=
-k_B\,
\text{tr}_S\{\hat{\rho}_S(t)\ln\hat{\rho}_S(t)\}
\geq 0
\quad,
\label{eq.entropy}
\end{equation} 
with Boltzmann constant $k_B$, which measures the entanglement between system and bath degrees of freedom\cite{Plenio07,Rivas10b,bouakline12,lorenz}. The equality in Eq.\eqref{eq.entropy} holds for a product state system-bath wave function, which is by definition non-entangled and here realized initially such that $S_{\text{vN}}(0)=0$. 

\begin{figure}[hbt]
\begin{center}
\includegraphics[scale=1.0]{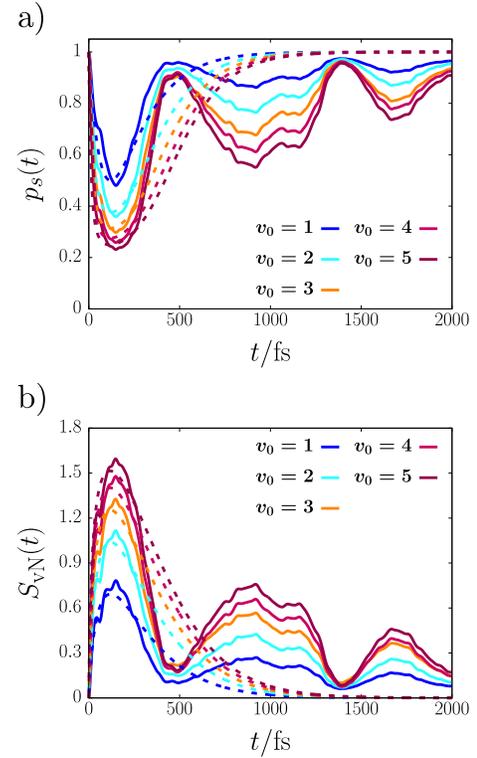}
\end{center}
\renewcommand{\baselinestretch}{1.}
\caption{Purity, $p_s(t)$, and von Neumann-entropy, $S_\mathrm{vN}(t)$, from full system-bath dynamics (bold) and Lindblad dynamics (dashed) as function of time for initial system states $\ket{v_0}$ with vibrational quantum numbers $v_0=1$ to $v_0=5$.}
\label{fig.entropy_purity}
\end{figure}

In Figs.\ref{fig.entropy_purity}(a) and (b), $p_S(t)$ and $S_{\text{vN}}(t)$ are shown for the relaxation of 
 various initial states 
 $|v_0\rangle$ up to $t_f=2000\,\mathrm{fs}$. For $t>0$, we find the system to be in a mixed state, due to finite system-bath entanglement, \textit{i.e.}, the full system-bath wave function, $\ket{\Psi_\mathrm{SB}(t)}$, is multiconfigurational in character and cannot be represented by a single product state. From the Markovian perspective, we find both the purity and von Neumann-entropy to tend against their limiting pure state values for times $t>1500\,\mathrm{fs}$, which indicates a complete population transfer to the system vibrational ground state. This observation is contrasted by the non-Markovian results, which (i) reflect the recurrences in the relaxation dynamics and (ii) do not lead to a pure system state due to non-vanishing system-bath entanglement in the time-interval shown. This observation is understood by referencing to Fig.\ref{fig.dynamics_mlmctdh} again, where we observed a number of damped recurrences in low-lying excited states and pronounced coherences between those states. In the long-time limit, however, a similar result can be expected for the full dynamics. Then the system ground state is the dominating system state in $\ket{\Psi_\mathrm{SB}(t)}$, while the bath contribution is given by a linear combination of excited bath states, \textit{i.e.}, system and bath state contributions factorize again in the long-time limit. 
 
We further see that the initial system state, $\ket{v_0}$, manifests in increasing oscillation amplitudes of both $p_S(t)$ and $S_{\text{vN}}(t)$ with increasing $v_0$. In passing we note that, for $v_0=1$, an analytical expression for the von Neumann-entropy can be given in the Lindblad case, namely\cite{bouakline12}
$
S_\mathrm{vN}(t) = k_B \left[\gamma_1 t \ e^{-\gamma_1 t} - \left(1-e^{-\gamma_1 t}\right) \text{ln}\left(1-e^{-\gamma_1 t}\right)
\right]$.
\subsubsection{System Energy Current}
\label{subsubsec.system_energy_current}
A final measure for (non-)Markovian behavior considered  in this work is the energy current between system and bath.
We discuss the dynamical details of the energy transfer process by examining the time-evolution of the system energy current, which we define as\cite{velizhanin2008}
\begin{equation}
\mathcal{J}_S(t)
=
\dfrac{\partial}{\partial t}\braket{\hat{H}_S}(t)
\begin{cases}
< 0, &\hspace{0.3cm} S\rightarrow B\\
> 0, &\hspace{0.3cm} S\leftarrow B
\end{cases}
\quad.
\label{eq.def_energy_current}
\end{equation}
The energy current captures both the magnitude and the direction of the energy transfer between system and bath, where we use the convention $\mathcal{J}_S(t)<0$ for energy flowing from the system to the bath ($S\rightarrow B$) and $\mathcal{J}_S(t)>0$ for energy back-flow ($S\leftarrow B$), respectively. The non-Markovian current is given by
\begin{equation}
\mathcal{J}_S(t)
=
\sum^{N_S-1}_{v=0}
\varepsilon_v\,
\dfrac{\partial \rho_{vv}(t)}{\partial t}
\quad,
\label{eq.full_system_energy_current}
\end{equation}
which takes the following form in the Lindblad formalism
\begin{equation}
\mathcal{J}_L(t)
=
-
\displaystyle\sum^{N_S-1}_{v=1}\,
\gamma_v\,
\Delta\varepsilon_v\,
\rho_{vv}(t)
\quad,
\label{eq.lindblad_system_energy_current}
\end{equation}
with $\Delta\varepsilon_v=\varepsilon_{v}-\varepsilon_{v-1}$. Detailed derivations are provided in Appendix C. In Fig.\ref{fig.current}, the time-evolution of both currents is shown for different initial system states.

\begin{figure}[hbt]
\begin{center}
\includegraphics[scale=1.0]{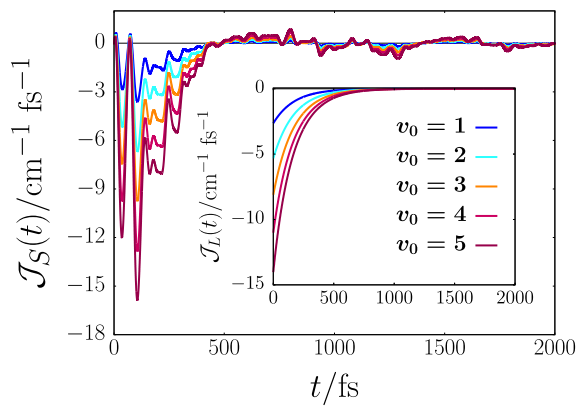}
\end{center}
\renewcommand{\baselinestretch}{1.}
\caption{System energy current from full system-bath dynamics, $\mathcal{J}_S(t)$, and Lindblad dynamics, $\mathcal{J}_L(t)$, (inset) as function of time for initial system states $\ket{v_0}$ with vibrational quantum numbers $v_0=1$ to $v_0=5$.}
\label{fig.current}
\end{figure}

At $t=0$, the current is negative in both approaches.  The magnitude of the currents is significantly larger for the Markovian case, which directly relates to the exponential system decay observed in the Lindblad dynamics, Fig.\ref{fig.dynamics_lindblad_vs_mlmctdh}a)-e), and to shorter half-lifetimes for the Lindblad case. Further, the Markovian current tends monotonically to zero reaching this limit for $t>1500\,\mathrm{fs}$, where the Markovian relaxation process is completed according to the population dynamics discussed above. In contrast, the time-evolution of the non-Markovian current is highly non-monotonic in nature and the energy-exchange occurs in a ``burst''-like fashion, which we qualitatively relate to the non-trivial character of the bath vDOS and the related system-bath coupling coefficients.

Further, $\mathcal{J}_L(t)\leq 0$ for all times $t$, \textit{i.e.}, the Markovian energy current is exclusively unidirectional from the system to the bath. In contrast, the full current is found to be positive for some time intervals, due to an energy back-flow from the bath to the system. Notably, this is not only the case for short times, where we found the post-rotating wave effects to be relevant, but also for times around $700\,\mathrm{fs}$ and $1600\,\mathrm{fs}$, \textit{i.e.}, around the onset of the recurrences in the non-Markovian population, where the non-Markovian current oscillates around a mean value of zero. Analogously to purity and von Neumann-entropy, the energy current's amplitude increases with the initial state vibrational quantum number, $v_0$. 
\section{Conclusion and Outlook}
\label{sec.conclusion}
In summary, in this work we have first of all extended a previous HEM-ML-MCTDH treatment of the vibrational relaxation of D-Si-Si bending mode due to vibration-phonon coupling at $T=0$ K\cite{fischer2020}, to higher initial states, $v_0=3$, $4$ and $5$. The efficiency of the HEM model in combination with ML-MCTDH allowed us to do so, avoiding thus the ``curse of dimensionality'' issue, which would otherwise make this problem with $>2000$ bath modes intractable, because the effort scales roughly as  $\sim P^{v_0}$. Here, $P=N_B$ if no HEM representation is used, with $N_B>2000$, and $P=M$ if the HEM is employed, with $M=60$ effective modes. With this method, we found that the excited state lifetimes decay faster with increasing $v_0$, however, not according to an ideal scaling law, $T_{1/2}^{(v_0)} \propto v_0^{-1}$, despite the system being rather harmonic in our example. 
The vibrational density of states and the coupling functions, however, are non-trivial in our case. This in turn causes non-trivial behavior of state populations, $\rho_{vv}(t)$, including damped oscillations and recurrences.

While the HEM-ML-MCTDH is efficient and very general, it is not always simple to apply and it still comes with non-negligible numerical effort. This can be an issue if, for example, longer propagation times are needed to describe a slower physical process which in turn requires $M$ to become larger. Therefore, as a second aspect of this work, we also applied the so-called multi-Davydov-D2 method to solve the system-bath TDSE in HEM representation. This method turns out, for the present form of the system-bath Hamiltonian at least, to perform very well in terms of computation time. Moreover, it can be systematically converged w.r.t. to basis size like (ML-)MCTDH, but with only a single control parameter, the ``multiplicity'', $K$. It remains to be seen if the method performs equally well for more complicated system-bath problems, {\it e.g.}, coupled and / or anharmonic baths. 
 
Finally, as a third point, with the ``exact'' system-bath dynamics for a non-trivial problem at hand, we were able to compare them to more approximate, reduced approaches such as Lindblad open-system density matrix theory. The Markovian approximation could be tested in this way, for a concrete and realistic example. Using various measures for non-Markovianity and / or additional approximations made in the Lindblad model, we find that both at short {\it and} at long times deviations between the two exist. The deviations can be seen, for the ``exact model'', in oscillatory and non-smooth behavior of vibrational populations, purity of the reduced density matrix, the von Neumann-entropy and energy flow between system and bath. For the latter, one in fact finds some back-flow of energy from bath to system, in contrast to what Lindblad theory would predict. Further, the exact solution of the TDSE shows the occurrence of non-vanishing coherences, in contrast to the Lindblad model.

To conclude, we have tested efficient numerical tools / new combinations of tools to study system-bath dynamics in non-trivial, realistic cases as they emerge, for example, in surface science. Possible next lines of research are to extend this work to finite temperature, to multi-dimensional systems, more complicated system-bath couplings (including multi-phonon processes), and systems or baths driven by external radiation.
\section*{Acknowledgements}
EWF and PS thank the Deutsche Forschungsgemeinschaft (DFG) for financial support through project Sa 547/18-1. {FG is also grateful for financial support by the DFG through project GR 1210/8-1.} EWF acknowledges support by the International Max Planck Research School for Elementary Processes in Physical Chemistry (IMPRS-EPPC) of the Fritz-Haber-Institute, Berlin. {MW acknowledges support by the International Max Planck Research School for Many Particle Systems
in Structured Environments (IMPRS-MPSSE) of the Max-Planck-Institute for the Physics of Complex Systems, Dresden.}

\section*{Data Availability Statement}
The data that support the findings of this study are available from the corresponding author upon reasonable request.

\section*{Conflict of Interest}
The authors have no conflicts to disclose.

\newpage
\section*{Appendix}
\subsection{Numerical Details of ML-MCTDH}
Here we provide numerical details of the ML-MCTDH calculations performed in this work, with the Heidelberg MCTDH package\cite{heidelbergmctdh}. For the $M^\mathrm{th}$-order truncated HEM Hamiltonian Eq.\eqref{eq.hem_hamiltonian}, we employed a spectral representation for the system in terms of its respective eigenstates and treated the effective mode bath in second quantization representation (SQR). We realized the Hamiltonian in the Heidelberg MCTDH package {\it via} invoking the analogy to non-adiabatic dynamics in vibronic coupling theory. Accordingly, $N_S=10$ vibrational system eigenstates formally take the role of electronic states with $n_s=10$ SPFs in the ML-wave function.

\begin{figure}[hbt]
\begin{center}
\includegraphics[scale=1.0]{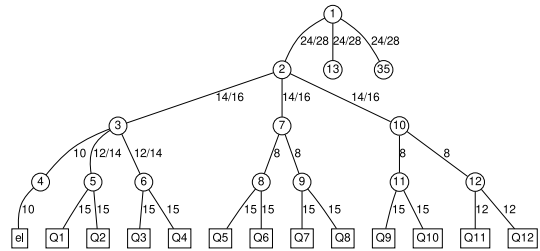}
\end{center}
\renewcommand{\baselinestretch}{1.}
\caption{First branch of the Multilayer-tree involving the system mode and the twelve effective modes as employed for vibrational relaxation dynamics with initial system states $\ket{v_0}$ for vibrational quantum numbers $v_0=1$ to $v_0=5$. Numbers next to edges for all but the lowest level correspond to SPFs employed where smaller number holds for $v_0=1,\dots,4$ and the higher number holds for $v_0=5$. Numbers next to edges connecting to the lowest layer correspond to primitive basis functions.}
\label{fig.mltree_branch_1}
\end{figure}
Further, the effective bath modes in SQR are conveniently represented on a sin-DVR grid starting at zero with a mesh width of one\cite{heidelbergmctdh}. Independent of the initial system state, we consider $N_\kappa=15$ primitive basis functions for the first ten effective modes and $N_\kappa=12$ grid points for all remaining modes.

\begin{figure*}[hbt]
\begin{center}
\includegraphics[scale=1.0]{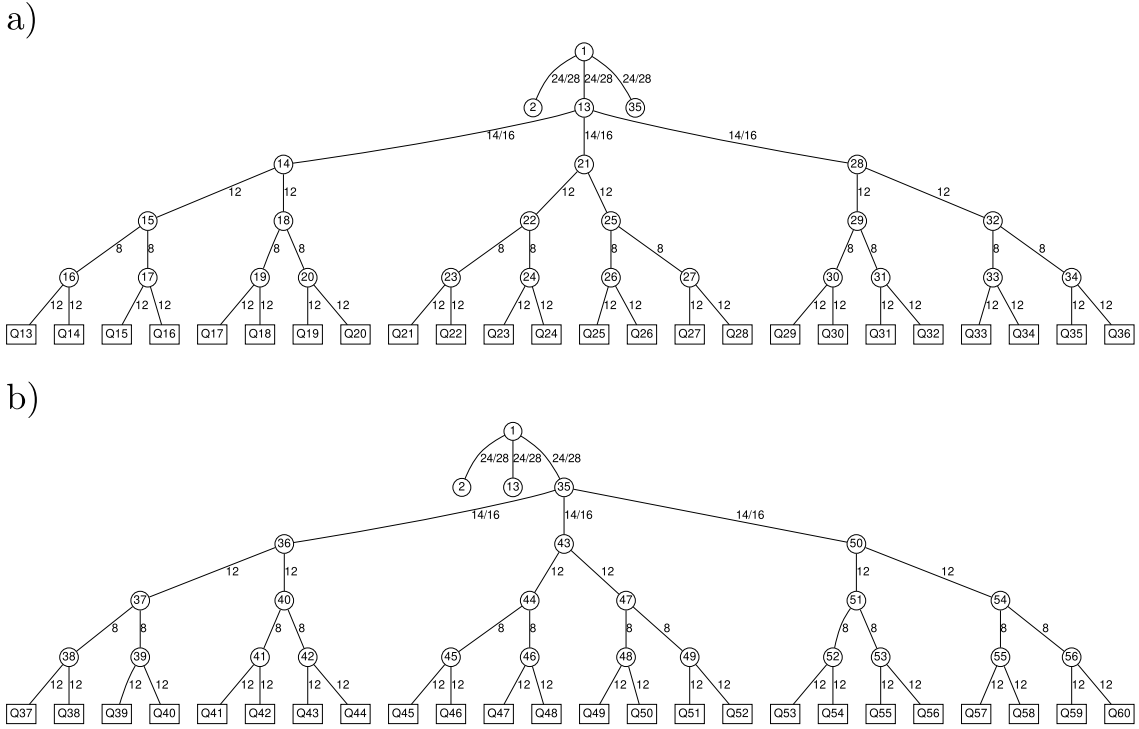}
\end{center}
\renewcommand{\baselinestretch}{1.}
\caption{Second and third branch of the Multilayer-tree involving effective modes $13$ to $36$ and $37$ to $60$ as employed for vibrational relaxation dynamics with initial system states $\ket{v_0}$ for vibrational quantum numbers $v_0=1$ to $v_0=5$. Numbers next to edges for all but the lowest level correspond to SPFs employed where smaller number holds for $v_0=1,\dots,4$ and the higher number holds for $v_0=5$. Numbers next to edges connecting to the lowest layer correspond to primitive basis functions.}
\label{fig.mltree_branch_2_3}
\end{figure*}

The structure of the ML-tree used in this work is shown in Figs.\ref{fig.mltree_branch_1} and \ref{fig.mltree_branch_2_3}. It is independent of the initial system state. We follow Appendix C of Ref.\cite{fischer2020} and divide the ML-tree into three branches corresponding to modes close to the system (1-12), an intermediate region (13-36) and a region ``far'' from the system (37-60). The number of SPFs employed to obtain natural populations $\leq 10^{-4}$ for all nodes in the ML-tree which we use here, does depend on the initial system state due to increasing correlation between the degrees of freedom with increasing $v_0$.
For number pairs of SPFs given in Figs.\ref{fig.mltree_branch_1} and \ref{fig.mltree_branch_2_3} next to the edges, the smaller ones relate to initial system quantum numbers $v_0=1,\dots,4$ and the larger ones to $v_0=5$, respectively.

{Equivalently, the $M^\mathrm{th}$-order truncated HEM Hamiltonian Eq.\eqref{eq.hem_hamiltonian} can be written in coordinate representation for both system and effective modes as discussed in Ref.\cite{fischer2020}. The same numerical results as discussed here are obtained by representing the system vibrational mode in terms of a harmonic oscillator (HO) DVR with $N_S=30$ grid points. The number of ``system''-SPFs is by $n_s=12$ for $v_0=1,\dots,3$ and $n_s=18$ SPFs for $v_0=4,5$, respectively. For the dimensionless effective bath modes, a HO-DVR is applied with $N_\kappa=15$ grid points for the first ten modes and $N_\kappa=12$ for the remaining ones. Here, the number of SPFs are equivalent to the ones presented in Figs.\ref{fig.mltree_branch_1} and \ref{fig.mltree_branch_2_3}.}
\subsection{Fermi's Golden Rule One-Phonon Relaxation Rates}
According to Fermi's Golden Rule, relaxation rates for one-phonon transitions in the open-system LvN-equation in Lindblad form \eqref{eq.liouville_von_Neumann_Lindblad} at $T=0$ K are given by\cite{lorenz2017,bouakline2019} 
\begin{equation}
\gamma_v
=
\pi\,
\vert
q_{vv-1}
\vert^2
\sum_{k=1}^{N_B}
\dfrac{c^2_k}{\omega_k}
\,
\delta(\Delta\varepsilon_{vv-1}-\hbar\omega_k)
\quad,
\label{eq.fermis_golden_rule}
\end{equation}
with coupling coefficients $c_k$ from Eq.(\ref{eq.system_bath_interaction}), harmonic bath frequencies $\omega_k$ and vibrational transition matrix elements
\begin{equation}
q_{vv-1}
=
\braket{v\vert q \vert v-1}
\quad .
\end{equation}
The Dirac delta function, $\delta(\Delta\varepsilon_{vv-1}-\hbar\omega_k)$, depends on both the energy difference, $\Delta\varepsilon_{vv-1}=\varepsilon_v-\varepsilon_{v-1}$, between two system eigenstates $\ket{v}$ and $\ket{v-1}$ and the harmonic bath frequencies. Here, we approximate $\delta(\Delta\varepsilon_{vv-1}-\hbar\omega_k)$ by a Lorentzian
\begin{equation}
\delta(\Delta\varepsilon_{vv-1}-\hbar\omega_k)
\approx
\dfrac{1}{\pi}
\dfrac{\sigma}{\sigma^2+(\Delta\varepsilon_{vv-1}-\hbar\omega_k)^2}
\quad,
\label{eq.dirac_lorentzian}
\end{equation}
where we chose a finite width of $\sigma=10\,\text{cm}^{-1}$, following Ref.\cite{bouakline2019}.

We note that in this work we consider only transitions $v \rightarrow v-1$ in the rate expressions, neglecting overtone transitions $v \rightarrow v-2, v-3 \dots$. These are possible, in principle, even for bilinear coupling $\propto q\,x_k$ (where $q$ is the system and $x_k$ are phonon mode), because the system Hamiltonian developed in Ref.\cite{lorenz2017} is slightly anharmonic. These overtone transitions rates were found to be several orders of magnitude slower than  the fundamental transitions and were therefore neglected, leading to the simplified LvN-Lindblad equation (\ref{eq.liouville_von_Neumann_Lindblad}). We also note that the fundamental rates $\gamma_v$ depend somewhat on the choice of the width parameter $\sigma$ in the delta function.

\subsection{Derivations of the System Energy Current}
\label{app.derivation_current_reduced_dynamics}
Here we derive expressions of the system energy current for the full system-bath dynamics and the reduced Markovian dynamics described by the Liouville-von Neumann equation in Lindblad form.

\subsubsection{Non-Markovian System Energy Current}
\label{subsubsec.current_full_dynamics}
The non-Markovian system energy current for the full system-bath quantum dynamics can be written,
 with the definition in Eq.(\ref{eq.def_energy_current}), as
\begin{equation}
\mathcal{J}_S(t)
=
\dfrac{\partial}{\partial t}
\text{tr}_S\{\hat{\rho}_S(t)\hat{H}_S\}
\quad,
\end{equation}
where the trace $\text{tr}_S\{\dots\}$ runs over the system energy eigenstates $\ket{v}$. 
 With the system Hamiltonian  $\hat{H}_S=\sum_v\varepsilon_v\ket{v}\bra{v}$ given 
 in (\ref{eq.zero_order_hamiltonian}), the trace is simply 
\begin{equation}
\mathcal{J}_S(t)
=
\sum^{N_S-1}_{v=0}
\varepsilon_v\,
\dfrac{\partial \rho_{vv}(t)}{\partial t}
\quad,
\label{eq.full_system_energy_current}
\end{equation}
where $\rho_{vv}(t)$ are the populations for the full non-Markovian system-bath dynamics and $\varepsilon_v$ are eigenenergies of the vibrational system, respectively. We note, Eq.\eqref{eq.full_system_energy_current} holds in principle for arbitrary populations obtained from both non-Markovian and Markovian approaches. 
\subsubsection{Markovian System Energy Current}
\label{subsubsec.current_reduced_dynamics}
The Markovian system energy current is derived from the Liouville-von Neumann equation in Lindblad form (\ref{eq.liouville_von_Neumann_Lindblad}). We rewrite the latter in compact form as 
\begin{equation}
\dfrac{\partial }{\partial t}
\hat{\rho}_S(t)
=
{{\mathcal{L}}}\,
\hat{\rho}_S(t)
\quad,
\label{eq.formal_solution_lindblad}
\end{equation} 
with Lindblad-Liouvillian, ${{\mathcal{L}}}={{\mathcal{L}}}_0+{{\mathcal{L}}}_D$, comprising the unitary part ${\cal{L}}_0$ and  ${\cal{L}}_D$ the dissipative part on the r.h.s of Eq.(\ref{eq.formal_solution_lindblad}). Following the same argument as in the last section, we find
\begin{equation}
\mathcal{J}_L(t)
=
\dfrac{\partial}{\partial t}
\text{tr}_S\{\hat{\rho}_S(t)\hat{H}_S\}
=
\text{tr}_S\{\dfrac{\partial}{\partial t}\hat{\rho}_S(t)\hat{H}_S\}
\quad ,
\end{equation}
as $\hat{H}_S$ is not explicitly time dependent.  which turns with Eq.\eqref{eq.formal_solution_lindblad} into
\begin{align}
\mathcal{J}_L(t)
&=
\text{tr}_S\{\hat{H}_S\,{{\mathcal{L}}}\,\hat{\rho}_S(t)\}
\quad.
\end{align}
Taking into account the explicit form of the Lindbladian, ${{\mathcal{L}}}$, and 
 considering that the unitary part is energy-conserving, we get
\begin{equation}
\mathcal{J}_L(t)
 = 
\text{tr}_S\{\hat{H}_S \sum_{v}\gamma_v
\left(\hat{C}_v\hat{\rho}_S(t) \hat{C}^{\dagger}_v
-\dfrac{1}{2}
[\hat{C}^{\dagger}_v \hat{C}_v \ \hat{\rho}_S(t)]_+
\right)\}
\quad.
\end{equation}
This gives 
\begin{align}
\mathcal{J}_L(t)
&=
\displaystyle\sum_{v=1}\,\gamma_v\,
\text{tr}_S\{\hat{H}_S\,\hat{C}_v\hat{\rho}_S(t)\hat{C}^{\dagger}_v\}
\nonumber
\\
&\hspace{0.2cm}
-
\displaystyle\sum_{v=1}\,\dfrac{\gamma_v}{2}\,
\text{tr}_S\{\hat{H}_S\,\hat{C}^{\dagger}_v \hat{C}_v\,\hat{\rho}_S(t)\}
\nonumber
\\
&\hspace{0.4cm}
-
\displaystyle\sum_{v=1}\,\dfrac{\gamma_v}{2}\,
\text{tr}_S\{\hat{H}_S\,\hat{\rho}_S(t)\hat{C}^{\dagger}_v \hat{C}_v\}
\quad,
\end{align}
which turns after performing the traces and employing, $\hat{H}_S=\sum_v\varepsilon_v\ket{v}\bra{v}$, into
\begin{align}
\mathcal{J}_L(t)
&=
\displaystyle\sum_{v=1}\,\gamma_v\,\varepsilon_{v-1}\,
\rho_{vv}(t)
\nonumber
\\
&\hspace{0.2cm}
-
\displaystyle\sum_{v=1}\,\dfrac{\gamma_v}{2}\,\varepsilon_{v}\,
\rho_{vv}(t)
\nonumber
\\
&\hspace{0.4cm}
-
\displaystyle\sum_{v=1}\,\dfrac{\gamma_v}{2}\,\varepsilon_{v}\,
\rho_{vv}(t)
\quad.
\end{align}
The desired result is finally obtained as
\begin{equation}
\mathcal{J}_L(t)
=
-
\displaystyle\sum_{v=1}\,
\gamma_v\,
\Delta\varepsilon_v\,
\rho_{vv}(t)
\quad,
\end{equation}
with $\Delta\varepsilon_v=\varepsilon_{v}-\varepsilon_{v-1}$ and properties
\begin{equation}
\mathcal{J}_L(t)
\leq 
0,\,\forall  t \geq 0
\quad,
\end{equation}
and
\begin{equation}
\mathcal{J}_L(0)
=
-
\gamma_{v_0}\,
\Delta\varepsilon_{v_0}
\quad.
\end{equation}
In the ``ideal scaling'' case (linear oscillator bilinearly coupled to a harmonic bath), we have $\gamma_{v_0}=\gamma_1 v_0$ and thus ($\rho_{v_0v_0}=1$), the initial energy current is $\mathcal{J}_L(0) \propto v_0$, which is roughly what can be seen from Fig.\ref{fig.current}. We finally note, $\mathcal{J}_L(t)$ is equivalently obtained by combining the equation of motion \eqref{lind1} with the general expression in Eq.\eqref{eq.full_system_energy_current}.


\begin{thebibliography}{100}

\bibitem{GuyotSionnest1995} P. Guyot-Sionnest, P. Lin, E. Miller, \textit{J. Chem. Phys} {\bf 102}, 4269 (1995).

\bibitem{huang2000} C.~Huang, C.~Rettner, D.~Auerbach, A.~Wodtke, \textit{Science} \textbf{290}, 111 (2000).

\bibitem{kroes2016} G.~J.~Kroes, C.~Diaz, \textit{Chem. Soc. Rev.} \textbf{45}, 3658 (2016).

\bibitem{guo1999} H.~Guo, P.~Saalfrank, T.~Seideman, \textit{Prog. Surf. Sci.} \textbf{62}, 239 (1999).

\bibitem{avouris} T. -C. Shen, C. Wang, G. Abeln, J. Tucker, J. Lyding, P. Avouris, R. Walkup, \textit{Science} {\bf 268}, 1590 (1995).

\bibitem{grasser} M. Jech, A.-M. El-Sayed, S. Tyaginov, D. Waldh\"or, F. Bouakline, P. Saalfrank,
          D. Jabs, Ch. Jungemann, M. Waltl, T. Grasser, \textit{Phys. Rev. Appl.} {\bf 16}, 014026 (2021).

\bibitem{saalfrank2006} P.~Saalfrank, \textit{Chem. Rev.} \textbf{106}, 4116 (2006).

\bibitem{arnolds2011} H.~Arnolds, \textit{Prog. Surf. Sci.} \textbf{86}, 1 (2011).

\bibitem{andrianov2006} I.~Andrianov, P.~Saalfrank, \textit{J. Chem. Phys.} \textbf{124}, 034710 (2006).

\bibitem{bouakline2017} F. Bouakline, U. Lorenz, G. Melani, G. K. Paramonov, P. Saalfrank, \textit{J. Chem. Phys.} \textbf{147}, 144703 (2017).

\bibitem{lorenz2017} U. Lorenz, P. Saalfrank, \textit{Chem. Phys.} \textbf{482}, 69 (2017).

\bibitem{bouakline2019} F. Bouakline, E. W. Fischer, P. Saalfrank, \textit{J. Chem. Phys.} \textbf{150}, 244105 (2019).

\bibitem{fischer2020} E. W. Fischer, F. Bouakline, M. Werther, P. Saalfrank, \textit{J. Chem. Phys.} \textbf{153}, 064704 (2020).


\bibitem{bixon1} M. Bixon, J. Jortner, \textit{J. Chem. Phys.} {\bf 48}, 715 (1968).

\bibitem{bixon2} A. Nitzan, J. Jortner, P. M. Rentzepis, \textit{Proc. R. Soc. A} {\bf 327}, 367 (1972).


\bibitem{tier1} E. Sibert III, W. Reinhardt, J.Hynes, \textit{J. Chem. Phys.} {\bf 81}, 1115 (1984).

\bibitem{tier2} K. Marshall, J. Hutchinson, \textit{J. Chem. Phys.} {\bf 91}, 3219 (1987).

\bibitem{tier3} A. Stuchebrukhov, R. Marcus, \textit{J. Chem. Phys.} {\bf 98}, 6044 (1993).

\bibitem{gindensperger2006} E. Gindensperger, I. Burghardt, L. S. Cederbaum, \textit{J. Chem. Phys.} \textbf{124}, 144103 (2006).

\bibitem{gindensperger2007a} E. Gindensperger, L. S. Cederbaum, \textit{J. Chem. Phys.} \textbf{127}, 124107 (2007).

\bibitem{gindensperger2007b} E. Gindensperger, H. K\"oppel, L. S. Cederbaum, \textit{J. Chem. Phys.} \textbf{126}, 034106 (2007).

\bibitem{hughes2009a} K. H. Hughes, C. D. Christ, I. Burghardt, \textit{J. Chem. Phys.} \textbf{131}, 024109 (2009).

\bibitem{hughes2009b} K. H. Hughes, C. D. Christ, I. Burghardt, \textit{J. Chem. Phys.} \textbf{131}, 124108 (2009).

\bibitem{burghardt2012} I. Burghardt, R. Martinazzo, K. H. Hughes, \textit{J. Chem. Phys.} \textbf{137}, 144107 (2012).

\bibitem{meyer1990} H.-D.~Meyer, U.~Manthe, L.~S.~Cederbaum, \textit{Chem. Phys. Lett.} \textbf{165}, 73 (1990).

\bibitem{manthe1992} U.~Manthe, H.-D.~Meyer, L.~S.~Cederbaum, \textit{J. Chem. Phys.} \textbf{97}, 3199 (1992).

\bibitem{beck2000} M.~H.~Beck, A.~J\"ackle, G.~A.~Worth, H.-D.~Meyer, \textit{Phys. Rep.} \textbf{324}, 1 (2000).

\bibitem{wang2003} H.~Wang, M.~Thoss, \textit{J. Chem. Phys.} \textbf{119}, 1289 (2003).

\bibitem{manthe2008} U.~Manthe, \textit{J. Chem. Phys} \textbf{128}, 164116 (2008).

\bibitem{vendrell2011} O.~Vendrell, H.-D.~Meyer, \textit{J. Chem. Phys} \textbf{134}, 044135 (2011).

\bibitem{wang2015} H.~Wang, \textit{J. Phys. Chem. A} \textbf{119}, 7951 (2015).


\bibitem{zhou2015} N.~Zhou, L.~Chen, D.~Xu, V.~Chernyak, Y.~Zhao, \textit{Phys.
Rev. B} \textbf{91}, 195129 (2015).

\bibitem{wang2016} L.~Wang , L.~Chen, N.~Zhou, Y.~Zhao, \textit{J. Chem. Phys.} \textbf{144}, 024101 (2016).

\bibitem{huang2017} Z.~Huang, L.~Wang, C.~Wu, L.~Chen, F.~Grossmann, Y.~Zhao, \textit{Phys. Chem. Chem. Phys.} \textbf{19}, 1655 (2017).

\bibitem{hartmann2019} R.~Hartmann, M.~Werther, F.~Grossmann, W.~T.~Strunz \textit{J. Chem. Phys.} \textbf{150}, 234105 (2019).

\bibitem{chen2019} L.~Chen, M.~F.~Gelin, W.~Domcke, \textit{J. Chem. Phys.} \textbf{150}, 024101 (2019).

\bibitem{paramonov2007a} G.~K.~Paramonov, I.~Andrianov, P.~Saalfrank, \textit{J. Phys. Chem. C} \textbf{111}, 5432 (2007).

\bibitem{paramonov2007b} G.~K.~Paramonov, S.~Beyvers, I.~Andrianov, P.~Saalfrank, \textit{Phys. Rev. B} \textbf{75}, 045405 (2007).

\bibitem{burghardt99}
I.~Burghardt, H.-D. Meyer, L.~Cederbaum, {\it J. Chem. Phys.} \textbf{111}, 2927 (1999).

\bibitem{burghardt03} I.~Burghardt, M.~Nest, G.~Worth, \textit{ J. Chem. Phys.} \textbf{119}, 5364 (2003).

\bibitem{martinazzo06} R. Martinazzo, M. Nest, P. Saalfrank, G-F. Tantardini, {\it J. Chem. Phys.} \textbf{125}, 194102 (2006).
 
\bibitem{shala2000} D.V. Shalashilin, M.S. Child, {\it J. Chem. Phys.} \textbf{113}, 10028 (2000).

\bibitem{child2003}
 M.S. Child, D.V. Shalashilin, {\it J. Chem. Phys.} \textbf{118}, 2061 (2003).

\bibitem{Sh09} D.~V. Shalashilin,  {\it J. Chem. Phys.}  \textbf{130}, 244101 (2009).

\bibitem{ShBu08}
D.~V. Shalashilin, I. Burghardt,  \textit{J. Chem. Phys.}  \textbf{129} (8), 084104 (2008).

\bibitem{irpc15}
G. W. Richings, I. Polyak, K. E. Spinlove, G. A. Worth, I. Burghardt, B. Lasorne, \textit{Int. Rev. Phys. Chem.}, \textbf{34},
269 (2015).

\bibitem{irpc21} M. Werther, S. Loho Choudhury, F. Grossmann, \textit{Int. Rev. Phys. Chem.} \textbf{40}, 81 (2021).

\bibitem{breuer2007} H. P. Breuer, F. Petruccione, \textit{The Theory of Open Quantum Systems} (Oxford
University Press, Oxford, 2007).

\bibitem{nitzan2014} A. Nitzan, \textit{Chemical Dynamics in Condensed Phases} (Oxford University Press, Oxford, 2014).

\bibitem{redfield1965} A.~G.~Redfield, \textit{Adv. Magn. Reson.} \textbf{1}, 1 (1965).


\bibitem{lindblad1976} G. Lindblad, \textit{Commun. Math. Phys.} \textbf{48}, 119 (1976).

\bibitem{gorini1976} V. Gorini, A. Kossakowski, E. C. G. Sudarshan, \textit{J. Math. Phys.} \textbf{17}, 821 (1976).

\bibitem{tanimura1989} Y.~Tanimura, R.~Kubo, \textit{J. Phys. Soc. Jpn.} \textbf{58}, 101 (1989).

\bibitem{tanimura2006} Y.~Tanimura, \textit{J. Phys. Soc. Jpn.} \textbf{75}, 082001 (2006).

\bibitem{tanimura2015} Y.~Tanimura, \textit{J. Chem. Phys.} \textbf{142}, 144110 (2015).

\bibitem{suess2014} D. Suess, A. Eisfeld, W. T. Strunz, \textit{Phys. Rev. Lett.} \textbf{113}, 150403 (2014).

\bibitem{tracenorm1} H.P.~Breuer, E.-M.~Laine, J.~Piilo, \textit{Phys. Rev. Lett.} \textbf{103}, 210401 (2009).

\bibitem{tracenorm2} H.P.~Breuer, \textit{J. Phys. B: At. Mol. Opt. Phys.} \textbf{45}, 154001 (2012).

\bibitem{lorenz} U. Lorenz, P. Saalfrank, \textit{Eur. Phys. J. D} \textbf{69}, 46 (2015).

\bibitem{liu2015} J. Liu, K. Sun, X. Wang, Y. Zhao, \textit{Phys. Chem. Chem. Phys.}, \textbf{17}, 8087 (2015).

\bibitem{thoss_spin_boson} S. Wenderoth, H.-P. Breuer,  M. Thoss, \textit{Phys. Rev. A} \textbf{104}, 012213 (2021).

\bibitem{Plenio07} M.B. Plenio, S. Virmani, \textit{Quant. Inf. Comput.} \textbf{7}, 1 (2007).

\bibitem{Rivas10b} {\'A.} Rivas, S.F. Huelga, M.B. Plenio, \textit{Phys. Rev. Lett.} \textbf{105}, 050403 (2010).

\bibitem{bouakline12} F. Bouakline, F. L{\"u}der, R. Martinazzo, P. Saalfrank, \textit{J. Phys. Chem. A} \textbf{116}, 11118 (2012).

\bibitem{werther2020a} M. Werther, F. Grossmann, \textit{Phys. Rev. A} \textbf{102}, 063710 (2020).

\bibitem{heidelbergmctdh} G. A. Worth, M. H. Beck, A. J\"ackle, and H.-D. Meyer, The MCTDH Package, Version 8.2, 2000; H.-D. Meyer, Version 8.3, 2002; Version 8.4, 2007; O. Vendrell and H.-D. Meyer, Version 8.5, 2013; Version. Version 8.5 contains the
ML-MCTDH algorithm. See http://mctdh.uni-hd.de. Used version: 8.6.1 (2021).

\bibitem{werther2020b} M. Werther, F. Grossmann, \textit{Phys. Rev. B} \textbf{101}, 174315 (2020).

\bibitem{zhao2021} Y. Zhao, K. Sun, L. Cheng, M. Gelin, \textit{WIREs Comput Mol Sci.} e1589 (2021).

\bibitem{johansson2012} J. R. Johansson, P. D. Nation, F. Nori, \textit{Comp. Phys. Comm.} \textbf{183}, 1760 (2012).

\bibitem{johansson2013} J. R. Johansson, P. D. Nation, F. Nori, \textit{Comp. Phys. Comm.} \textbf{184}, 1234 (2013).

\bibitem{fujihashi2017} Y. Fujihashi, L. Wang, Y. Zhao, \textit{J. Chem. Phys.} \textbf{147}, 234107, (2017).

\bibitem{wang2008} H. Wang, M. Thoss, \textit{New J. Phys.} \textbf{10}, 115005 (2008).

\bibitem{velizhanin2008} K. A. Velizhanin, H. Wang, M. Thoss, \textit{Chem. Phys. Lett.} \textbf{460}, 325 (2008).

\end{thebibliography}
\end{document}